\AtBeginDocument{\bibstyle{mnras}}
\documentclass[fleqn,usenatbib]{mn2e}

\usepackage{newtxtext,newtxmath}
\usepackage[T1]{fontenc}
\usepackage{ae,aecompl}
\usepackage{graphicx}	% Including figure files
\usepackage{amsmath}	% Advanced maths commands
\usepackage{amssymb}	% Extra maths symbols
\usepackage{times}
\usepackage{xcolor}
\usepackage{multicol}
\newcommand{\mrm}{\mathrm}  %shorter?
\newcommand{\aeff}{\alpha_{\mathrm{eff}}}
\newcommand{\abareff}{\bar{\alpha}_{\mathrm{eff}}}

\newcommand{\apjs}{ApJS}
\newcommand{\aap}{A\&A}
\newcommand{\apjl}{ApJL}

\title[Convection in Common Envelopes]{The Role of Convection in Determining the Ejection Efficiency of Common Envelope Interactions}

\author[Wilson \& Nordhaus]{E. C. Wilson$^{1}$\thanks{E-mail: ecw7497@rit.edu} and J. Nordhaus$^{1,\hspace{0.03cm}2}$\thanks{E-mail: nordhaus@astro.rit.edu}
\\
$^{1}$Center for Computational Relativity and Gravitation, Rochester Institute of Technology, NY 14623, USA\\
$^{2}$National Technical Institute for the Deaf, Rochester Institute of Technology, NY 14623, USA\\
}

\date{Accepted XXX. Received YYY; in original form ZZZ}

\pubyear{2018}

\begin{document}
\label{firstpage}
\pagerange{\pageref{firstpage}--\pageref{lastpage}}
\maketitle

\begin{abstract}
A widely used method for parameterizing the outcomes of common envelopes (CEs) involves defining an ejection efficiency, $\abareff$, that represents the fraction of orbital energy used to unbind the envelope as the orbit decays.  Given $\abareff$, a prediction for the post-CE orbital separation is possible with knowledge of the energy required to unbind the primary's envelope from its core.  Unfortunately, placing observational constraints on $\abareff$ is challenging as it requires knowledge of the primary's structure at the onset of the common envelope phase.  Numerical simulations have also had difficulties reproducing post-CE orbital configurations as they leave extended, but still bound, envelopes.  Using detailed stellar interior profiles, we calculate $\abareff$ values for a matrix of primary-companion mass pairs when the primary is at maximal extent in its evolution.  We find that the ejection efficiency is most sensitive to the properties of the surface-contact convective region (SCCR).   In this region, the convective transport timescales are often short compared to orbital decay timescales, thereby allowing the star to effectively radiate orbital energy and thus lower $\abareff$.  The inclusion of convection in numerical simulations of CEs may aid ejection without the need for additional energy sources as the orbit must shrink substantially further before the requisite energy can be tapped to drive ejection. Additionally, convection leads to predicted post-CE orbital periods of less than a day in many cases, an observational result that has been difficult to reproduce in population studies where $\abareff$ is taken to be constant.  Finally, we provide a simple method to calculate $\abareff$ if the properties of the SCCR are known.
\end{abstract}


%%%%%%%%%%%%%%%%%%%%%%%%%%%%%%%%%%%%%%%%%%%%%%%%%%

%%%%%%%%%%%%%%%%% BODY OF PAPER %%%%%%%%%%%%%%%%%%

\section{Introduction}

Common envelopes (CEs) are events that often occur in binary systems when one component, the primary, evolves off the main-sequence \citep{Paczynski1976,Ivanova2013,Kochanek:2014aa}.  Significant radial expansion of the primary during post-main-sequence evolution can lead to direct engulfment of the companion, Roche Lobe overflow, and/or orbital decay via tidal dissipation \citep{Nordhaus2006,Nordhaus:2010aa,Chen:2017aa}.  The result is a binary system consisting of the primary's core and the companion embedded in a CE formed from the primary's envelope.

Once immersed inside a CE, the orbit of the system has been shown to decay rapidly ($\lesssim$$1-1000$ years) making direct detection difficult \citep{Ivanova:2013aa} and precursor emission signatures a promising means of identification \citep{MacLeod:2018aa}.   However, in wide triple systems, the orbital decay timescale may be significantly longer \citep{Michaely2018}. Common envelope phases are thought to be the primary mechanism for producing short-period binaries in the universe \citep{Toonen2013,Kruckow:2018aa,Canals:2018aa}, though not the only method \citep{Fabrycky:2007aa,Thompson:2011aa,Shappee:2013aa, Michaely2016}.

During inspiral, energy and angular momentum are transferred from the orbit to the envelope \citep{Iben1993}.  If sufficient to eject the CE, a tight binary emerges that contains at least one compact object.  If the envelope cannot be ejected, the companion is destroyed leaving a ``single'' star that nevertheless underwent a binary interaction such that its evolution may be modified \citep{Nordhaus:2011aa}.

A method for parameterizing the outcomes of common envelopes, and hence predictions for their progeny populations, involves defining an ejection efficiency, $\abareff$, based on energetic arguments.  This ``$\alpha$-formalism'' is broadly used in population synthesis studies and often defined in the following way:
\begin{equation}
	\abareff = E_{\mrm{bind}}/\Delta E_{\mrm{orb}},
	\label{eq:traditionalalpha}
\end{equation}
where $E_{\mrm{bind}}$ is the energy required to unbind the primary's envelope from its core and $\Delta E_{\mrm{orb}}$ is the orbital energy released during inspiral (for a detailed discussion see Sec~\ref{sec:energycalc}; \citealt{Tutukov1979, Iben1984, Webbink1984, Livio1988, DeMarco2011}).  Given $\abareff$, and knowledge of the primary's binding energy, the post-CE orbital separations can then be determined.  Because the transfer of energy to the envelope is not perfectly efficient, and because predictions for the progeny populations are highly sensitive to adopted values \citep{Claeys2014}, better constraints on $\abareff$, from either observations or theory, are topics of active research.  In particular, an improved understanding of the ejection efficiency as a function of binary parameters or internal CE structure is needed as $\abareff$ is often taken to be constant.

Observations of CE progenitors have allowed some estimates of $\abareff$, albeit with significant uncertainties. \citet{Zorotovic2010} identified over 50 systems that are likely progeny of CE evolution, and determined that most are consistent with $\abareff\simeq 0.2-0.3$.  This is in general agreement with \cite{Cojocaru:2017aa}, who performed population synthesis studies of Galactic white dwarf main-sequence binaries using data from the Sloan Digital Sky Survey (SDSS) Data Release 12. Also utilizing population synthesis techniques, \citet{Davis2010} determined that $\abareff>0.1$ reasonably describes all systems with late-type secondaries, but produced an overabundance of post-CE systems with orbital periods greater than a day. In a similar manner, \citet{Toonen2013} argue that the ejection efficiency must be low to explain the observed post-CE orbital period distribution present in the SDSS sample.

Additional studies that have tested the dependence of the ejection efficiencies on the mass ratio of the binary, $q=m_2/M_1$, have produced conflicting results.  \citet{DeMarco2011} find that the mass ratio is in anti-correlation with $\abareff$; an increased companion mass results in a decreased $\abareff$.  \citet{Zorotovic2011}, by way of orbital separation, find that the mass ratio is in fact in correlation with $\abareff$, attributing the increased ejection efficiency to the increased initial orbital energy of the more massive companion. We discuss these contradictory findings in relation to our results in Section~\ref{sec:massratio}.

Three-dimensional hydrodynamic simulations of common envelopes have been carried out in recent years by multiple groups using diverse codes and numerical techniques \citep{Ricker2012,Passy2012,Ohlmann2015,Chamandy:2018aa}.  While inspiral occurs rapidly, the envelope is pushed outward yet remains bound.  This failure to eject the CE has resulted in proposed solutions that include additional energy sources (recombination/accretion/jets), processes that operate on longer timescales, or ejection through dust-driven winds \citep{Soker:2015aa,Ivanova:2015aa,Kuruwita:2016aa,Glanz:2018aa,Sabach2017,Grichener2018a,Kashi:2018aa,Ivanova:2018aa,Soker:2018aa}.  While such effects may in fact prove necessary, it is first useful to consider the physical effects incorporated in simulations.  The energy budget of the CE interaction is set by the initial orbital energy.  As inspiral occurs, liberated orbital energy will be transferred to the CE unless it is lost via radiation. In this context, it is interesting to note that hydrodynamic simulations do not include radiation, and therefore provide incomplete analyses of the ejection efficiency. A full examination of the energy components in CE simulations and a robust discussion of why the envelope remains bound at large distances is needed, i.e.~\citet{Chamandy:2018ab}.

There have been several previous studies that investigate the effects of convection in conjunction with recombination energy.  \citet{Grichener2018a} in particular consider a common envelope in which the inspiraling companion deposits energy and the envelope expands.  The authors find that convection efficiently transports recombination energy to surface radiative regions where it is lost.  The energy transport time is on the order of months and shorter than dynamical timescales.  This is consistent with the results we present in this work.  In an earlier study, \citet{Sabach2017} argue that when helium recombines, energy transport by convection cannot be neglected.  While it may increase the luminosity of the event, it cannot be used to unbind the envelope.  Whether recombination energy can be tapped to drive ejection remains a subject of vigorous debate \citep{Ivanova:2018aa,Soker:2018aa}.

In this paper, we focus on the general effects of convection, internal structure, and mass ratio on ejection efficiencies.  Post-main sequence giants possess deep and vigorous convective envelopes which can carry energy to the surface where it can be lost via radiation, effectively lowering the ejection efficiency.  In regions where radiative losses do not occur, convection can redistribute energy, carrying it to parcels of gas that are not in the direct vicinity of the binary, thus aiding ejection.

In Section 2, we describe our methodology, stellar models and the physics in which we ground our analysis. Our data are presented in Section 3. In Section 4 we discuss our results in the context of observational and theoretical work, and present our conclusions in Section 5. 

\begin{figure*}
\begin{multicols}{2}
 \centering
 \begin{minipage}{0.48\textwidth}
   \includegraphics[width=\textwidth]{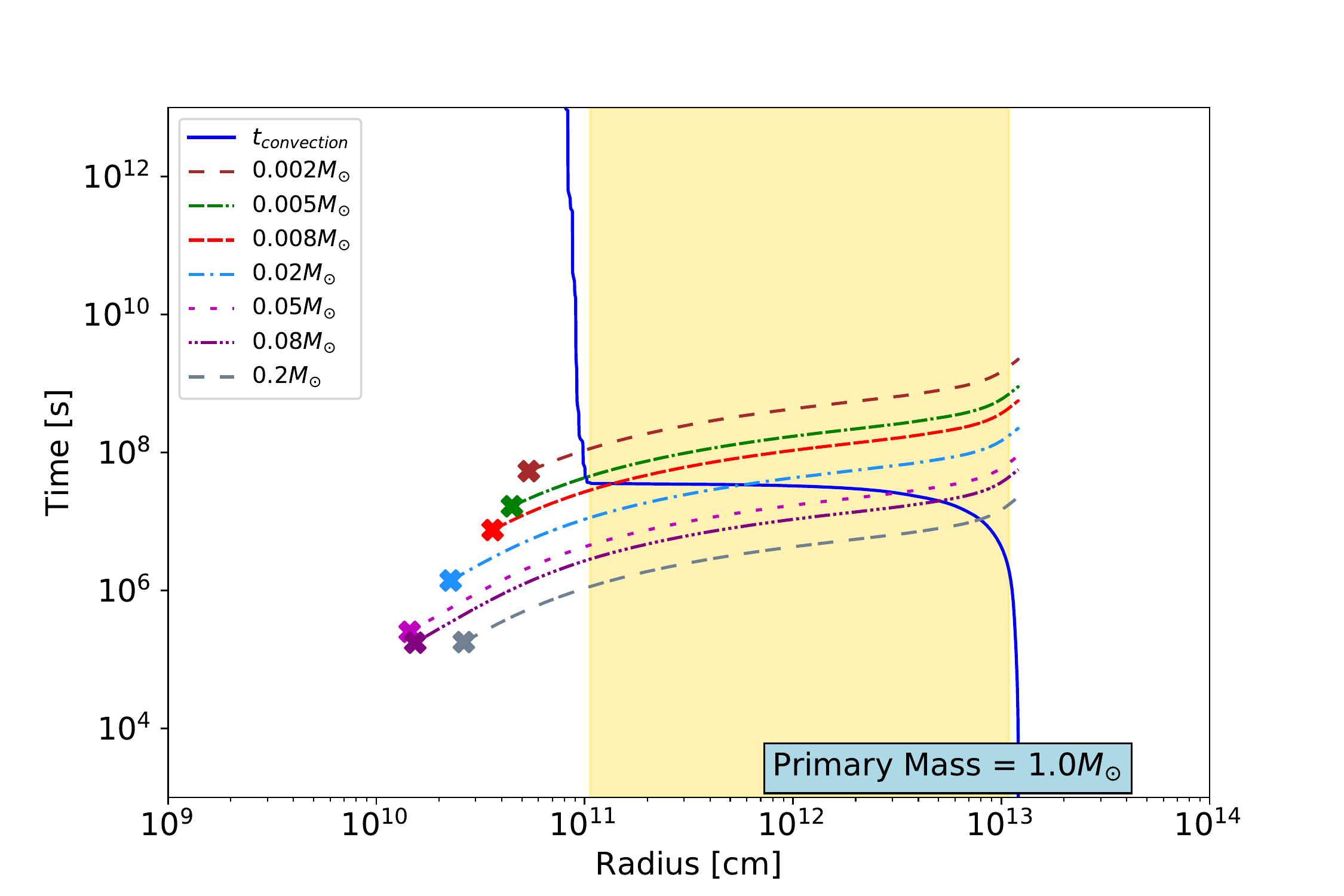}
 \end{minipage}
\hfill
 \begin{minipage}{0.48\textwidth}
   \includegraphics[width=\textwidth]{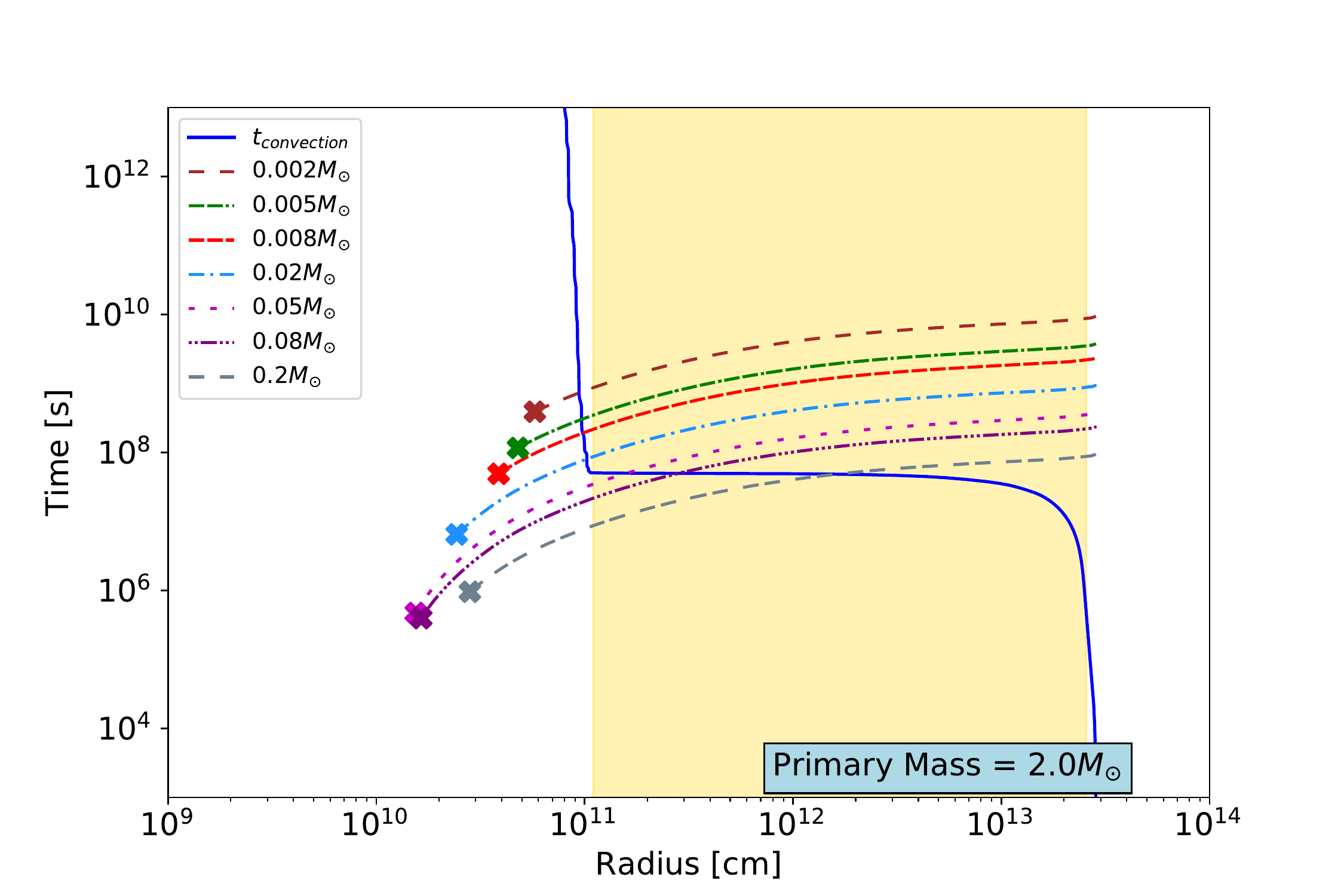}
 \end{minipage}
\hfill
 \begin{minipage}{0.48\textwidth}
   \includegraphics[width=\textwidth]{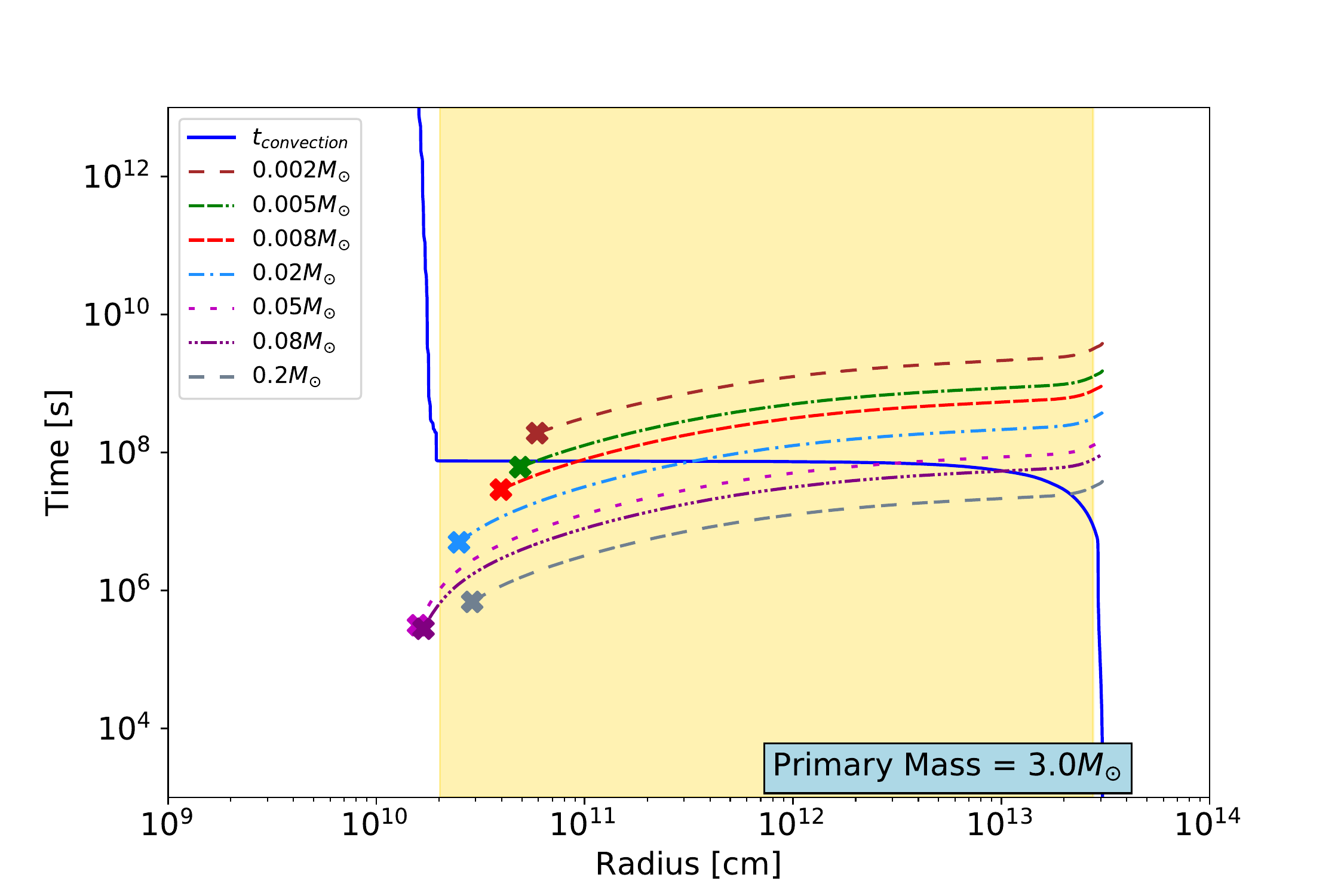}
 \end{minipage}
\hfill
 \begin{minipage}{0.48\textwidth}
   \includegraphics[width=\textwidth]{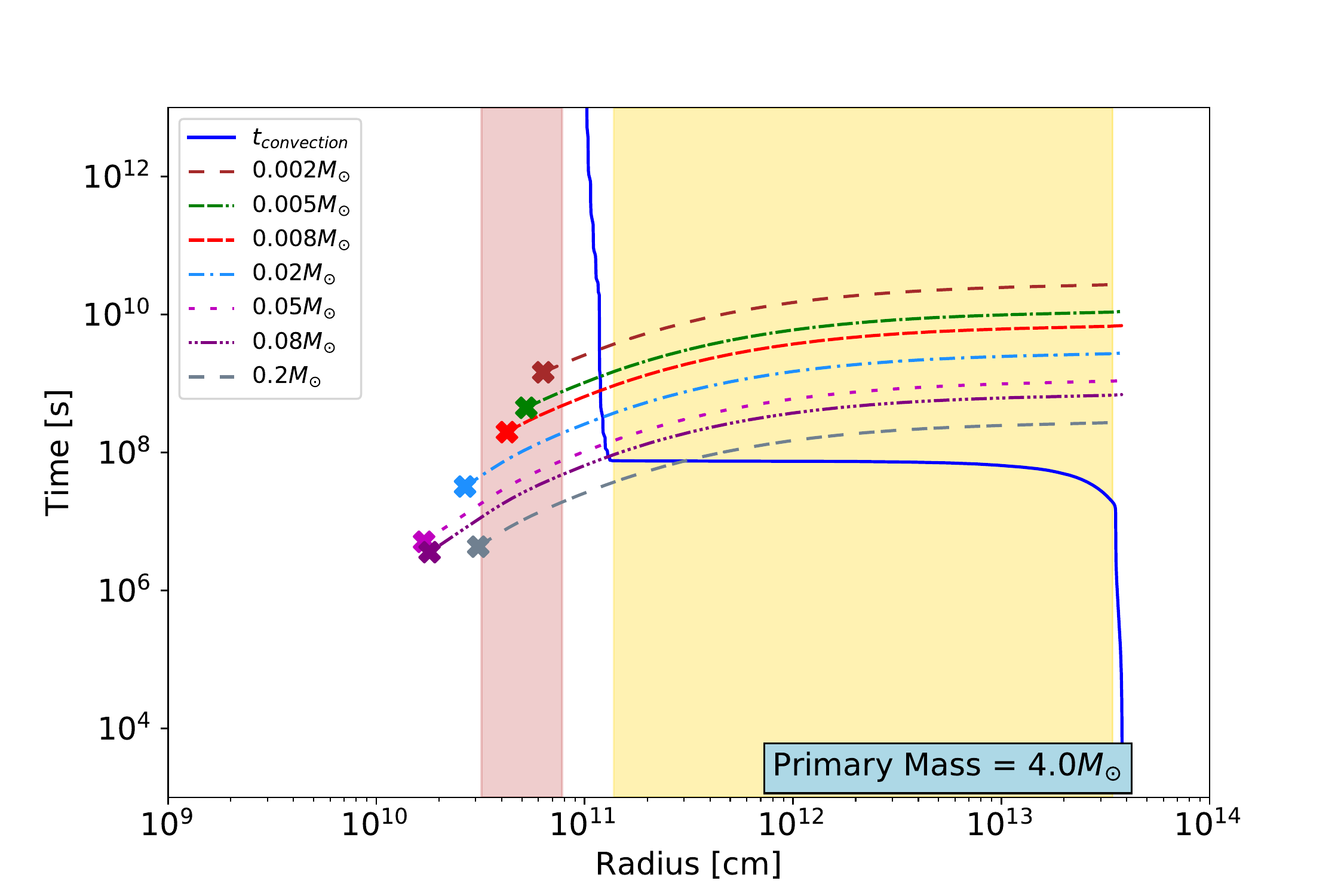}
 \end{minipage}
 \hfill
  \begin{minipage}{0.48\textwidth}
   \includegraphics[width=\textwidth]{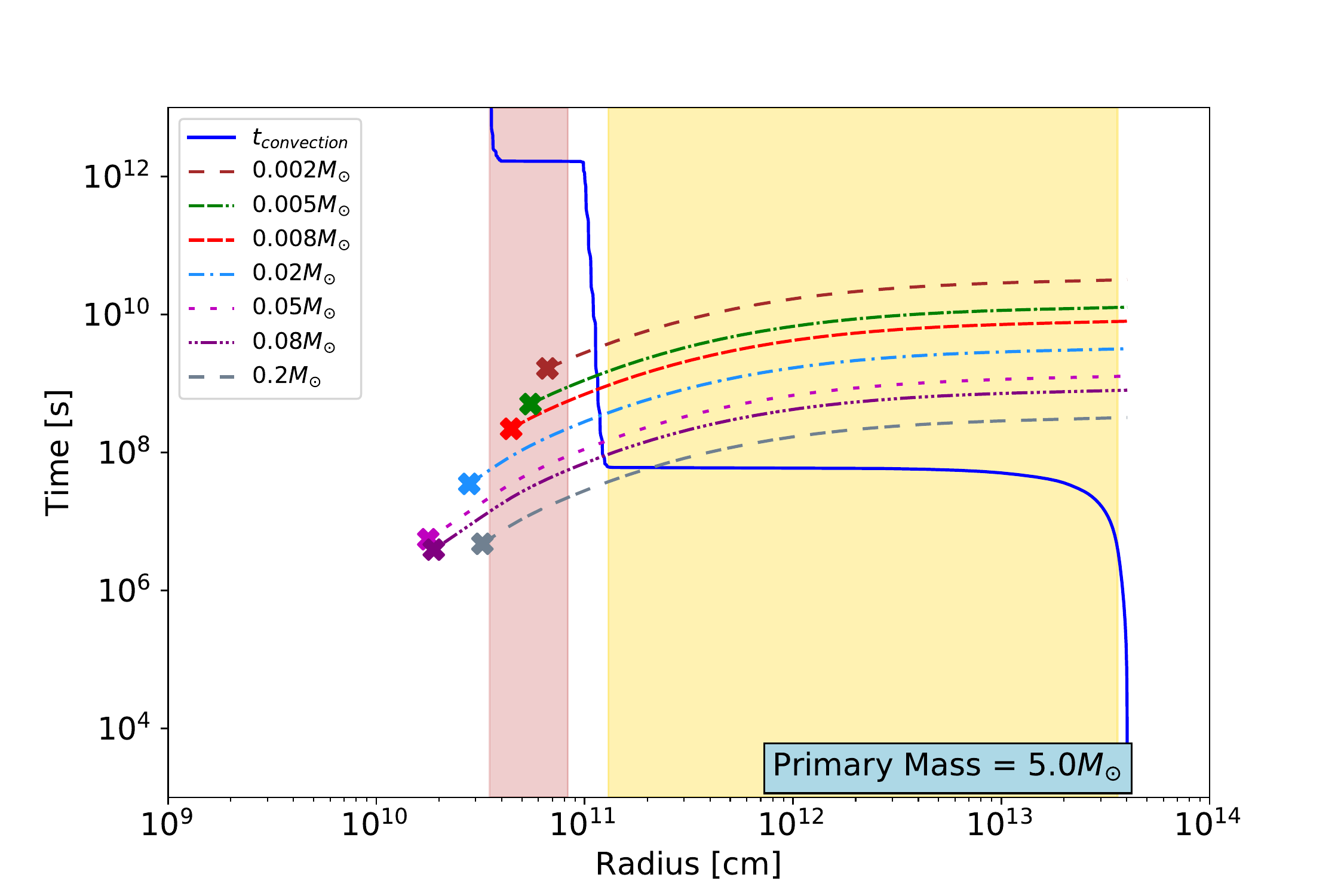}
 \end{minipage}
 \hfill
  \begin{minipage}{0.48\textwidth}
   \includegraphics[width=\textwidth]{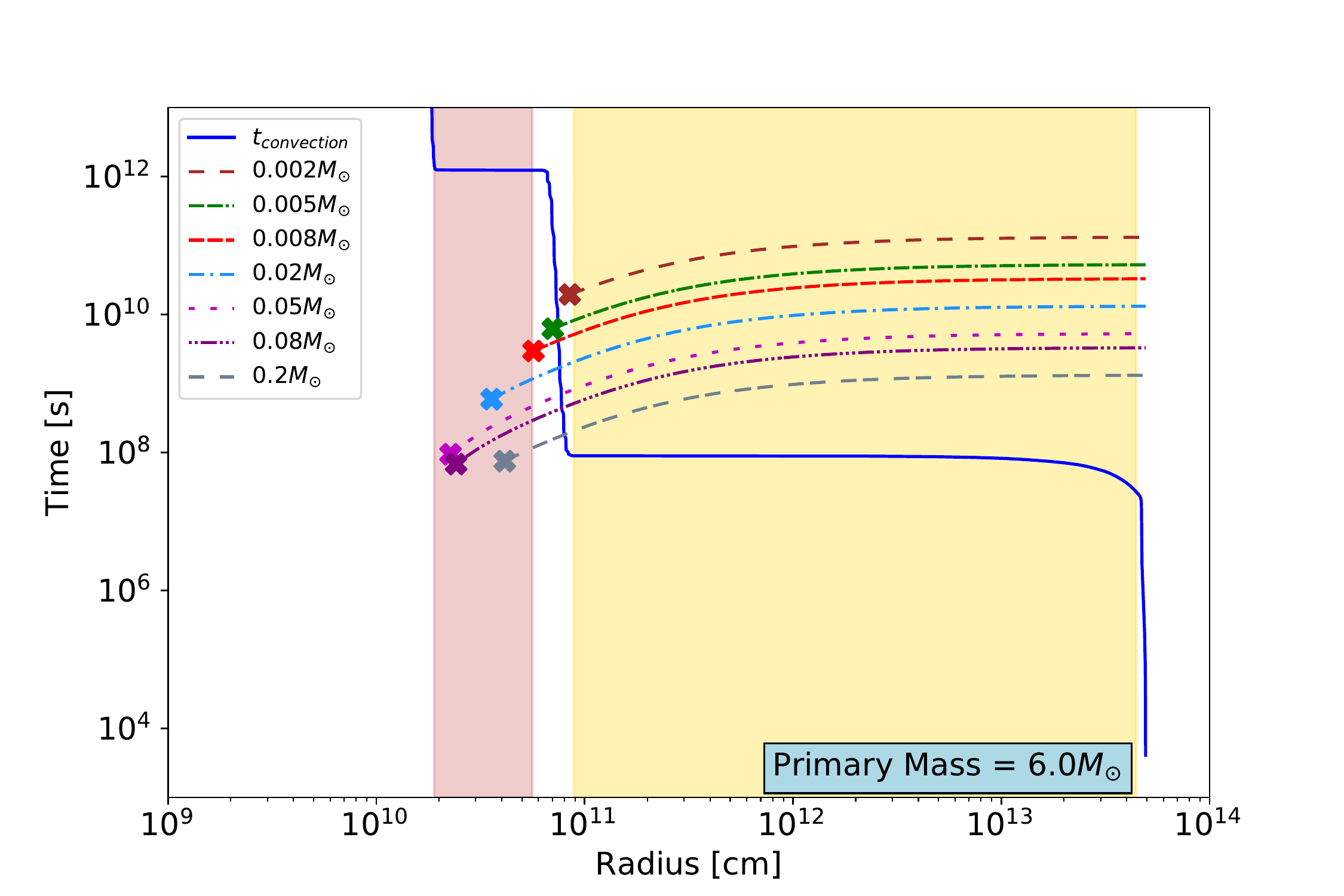}
 \end{minipage}
 \end{multicols}
 \hfill

 \caption{Comparative timescale plots for a sample of representative primary masses at their maximum radial extent and several test companion masses. The convective timescale profile of the primary star is shown in solid blue. The coloured, dashed lines show the inspiral timescale - the time it takes for the companion mass to spiral from its current radius to the centre of the primary star. The radius at which each companion mass shreds due to the gravity of the primary mass is marked with an X. The surface-contact convective regions (SCCRs) of the primary star that do not contribute to the unbinding of the envelope are shaded in yellow. Interior convective zones that do not extend to the primary's surface are shaded in pink.}
 \label{fig:Timescales}
\end{figure*}

\section{Methodology}
\label{sec:methodology}
Our stellar models were computed using \textsc{mesa} (release 10108), an open-source stellar evolution code that allows users to produce spherically symmetric models of stellar interiors during all phases of a star's evolution \citep{Paxton:2011aa, Paxton:2018aa}\footnote{\textsc{mesa} is available at http://mesa.sourceforge.net}.  Each star was evolved from the pre-main sequence to the white dwarf phase for zero-age-main-sequence (ZAMS) masses in the range of $0.8 M_{\odot} - 6.0 M_{\odot}$ in increments of $0.2M_\odot$ with finely-meshed time-stepping.  Stars with initial masses below $0.8 M_\odot$ were not included as they have not evolved off the main sequence during the lifetime of the universe.  Mass loss on the Red Giant Branch (RGB) followed a Reimer's prescription with $\eta_{\rm R}=0.7$ while mass loss on the Asymptotic Giant Branch (AGB) followed a Blocker prescription with $\eta_{\rm B}=0.7$ \citep{Reimers:1975aa, Bloecker:1995aa}.  All models were assumed to have solar metallicity (z=0.02).

For each evolutionary model, the interior profile at maximum extent was chosen as this is a likely time for a CE to occur since the primary occupies its greatest possible volume for engulfing companions.  This large volume and extended radius allow for strong tidal torques which shrink the companion's orbit \citep{Villaver:2009aa, Nordhaus:2010aa, Nordhaus:2013aa}.  Each interior profile contains radial information about the mass, density, convective properties, and core and envelope boundaries.  From these, we calculate the primary's binding energy, location of the convective zones, inspiral timescales, tidal disruption radii, and the energy released during orbital decay.  These quantities are used to determine $\abareff$ and the post-CE orbital separations for the companions that survive the CE interaction.

\subsection{Convective Regions}
\label{sec:convregions}
Post-main sequence stars host deep convective envelopes that can transport energy to optically thin surfaces where it is radiated away.  As the companion inspirals, there may be interior regions where convection can effectively carry newly liberated orbital energy to the surface.  The CE may then regulate itself with little-to-no orbital energy available for ejection until the companion reaches a region where the effects of convective transport no longer dominate.

To identify the convective regions of the primary star, we extract the calculated convective velocities ($v_{\mrm{conv}}$) from our interior profiles when each star is at the maximum radial extent in its evolution.  The convective timescale can then be found:
\begin{equation}
	t_{\mrm{conv}}[r] = \int_{r}^{R_{\star}}\frac{1}{v_{\mrm{conv}}} \mrm{d}r
	\label{eq:tconv}
\end{equation}
\citep{Grichener2018a}.
Similarly for each radius in the primary, we can determine the time required for the orbit to fully decay.  This inspiral timescale is given as,
\begin{equation}
	t_{\mrm{inspiral}}[r] = \int_{r_{\mrm{i}}}^{r_{\mrm{shred}}}{\frac{\left(\frac{\mathrm{d}M}{\mathrm{d}r}-\frac{M[r]}{r}\right)\ \sqrt[]	{v_r^2+\bar{v}_{\phi}^2}}{4 \xi \pi G m_2 r \rho[r]}\mrm{d}r}
	\label{eq:tinsp}
\end{equation}
where $r_i$ is the initial radial position, $r_{\mrm{shred}}$ is the tidal shredding radius which can be estimated via $r_{\mrm{shred}}\sim R_2 \sqrt[3]{2M_{\mrm{core}}/m_2}$, and $\bar{v}_{\phi}=v_{\phi}-v_{\mrm{env}} \simeq v_{\phi}$ for slow rotators such as RGB/AGB stars \citep{Nordhaus:2007aa}. The parameter $\xi$ accounts for the geometry of the companion's wake, the gaseous drag of the medium, and the Mach number \citep{Park2017}. We assume a value of $\xi=4$, and note that the ejection efficiency is not sensitive to this value for the mass ratios considered in this work.

For regions in which the $t_{\mathrm{conv}} {\ll} t_{\mrm{inspiral}}$, convection will transport orbital energy radially outward.  If the turbulent region reaches an optically thin area such as the surface of the star, this energy can be lost via radiation.  We refer to the convection region that makes contact with the surface of the star as the surface-contact convective region (SCCR).  For lower mass stars in our sample ($\lesssim 3.0 M_{\odot}$), there tends to be a single convective region at maximum extent. For stars more massive than $4.0 M_{\odot}$, a deeper, yet physically distinct, secondary convective layer is also present (see Figure~\ref{fig:Timescales}).

The inspiral and convective timescales are presented in Figure~\ref{fig:Timescales} for primary masses ranging from $1.0-6.0$ $M_\odot$ and companion masses ranging from $0.002-0.2$ $M_\odot$. SCCRs are shaded in yellow and secondary convective regions are shaded in pink. The tidal disruption locations are represented with X markers. The location and depth of the SCCR of the primary during inspiral is especially important and discussed in detail throughout the remainder of this paper.

\subsection{Energy and Luminosity Considerations}
\label{sec:energycalc}

The energy required to unbind the primary's envelope must be known to compute $\abareff$.  By carrying out calculations directly from our stellar evolution models, we avoid employing $\lambda$-formalisms, which approximate the primary's gravitational binding energy for situations in which the interior structure is not known \citep{DeMarco2011}.
The minimum energy required to strip the envelope's mass exterior to a radius, $r$, is then given by:
\begin{equation}
    E_{\mathrm{bind}}[r]=-\int_M^{M_\mrm{o}}\frac{G M[r]}{r}\mrm{d}m[r],
	\label{eq:Ebind}
\end{equation}
where $M_{\mrm{o}}$ is the total mass of the primary star.  One necessary, but not exclusive, condition for CE ejection is that the orbital energy released during inspiral must exceed the binding energy of the envelope.  Note that we focus exclusively on the gravitational component of the binding energy and do not include the internal energy of each shell in our calculations.  While this could affect the ejection efficiency of the system, it has not been shown to make a significant contribution to the binding energy \citep{Han:1995aa,Ivanova2013}.

The energy released via inspiral is given by
\begin{equation}
    \Delta E_{\mrm{orb}}[r]
    =\frac{G m_2}{2} \left(\frac{M[r_\mrm{i}]}{r_\mrm{i}}-\frac{M[r]}{r}\right)
	\label{eq:Eorb}
\end{equation}
where $m_2$ is the companion's mass and $r_\mrm{i}$ is the radius of the companion's orbit at the onset of energy transfer due to inspiral through the primary.

Equations~\ref{eq:Ebind} and~\ref{eq:Eorb} can then be combined with an efficiency to yield the following:
\begin{equation}
E_{\mathrm{bind}} = \abareff \Delta E_{\mathrm{orb}},
\label{eq:alpha}
\end{equation}
where $\abareff$ is the effective efficiency of energy transfer to the envelope from the decaying orbit.  If $\abareff=1$, then all transferred orbital energy remains in the system and can fully contribute toward ejection.  If $\abareff=0$, no orbital energy remains in the system and the CE would never be ejected.  We discuss $\abareff$ in the context of models and their convective zones in detail in Section~\ref{sec:transfer}.

\begin{figure}
\centering
\includegraphics[width=0.5\textwidth]{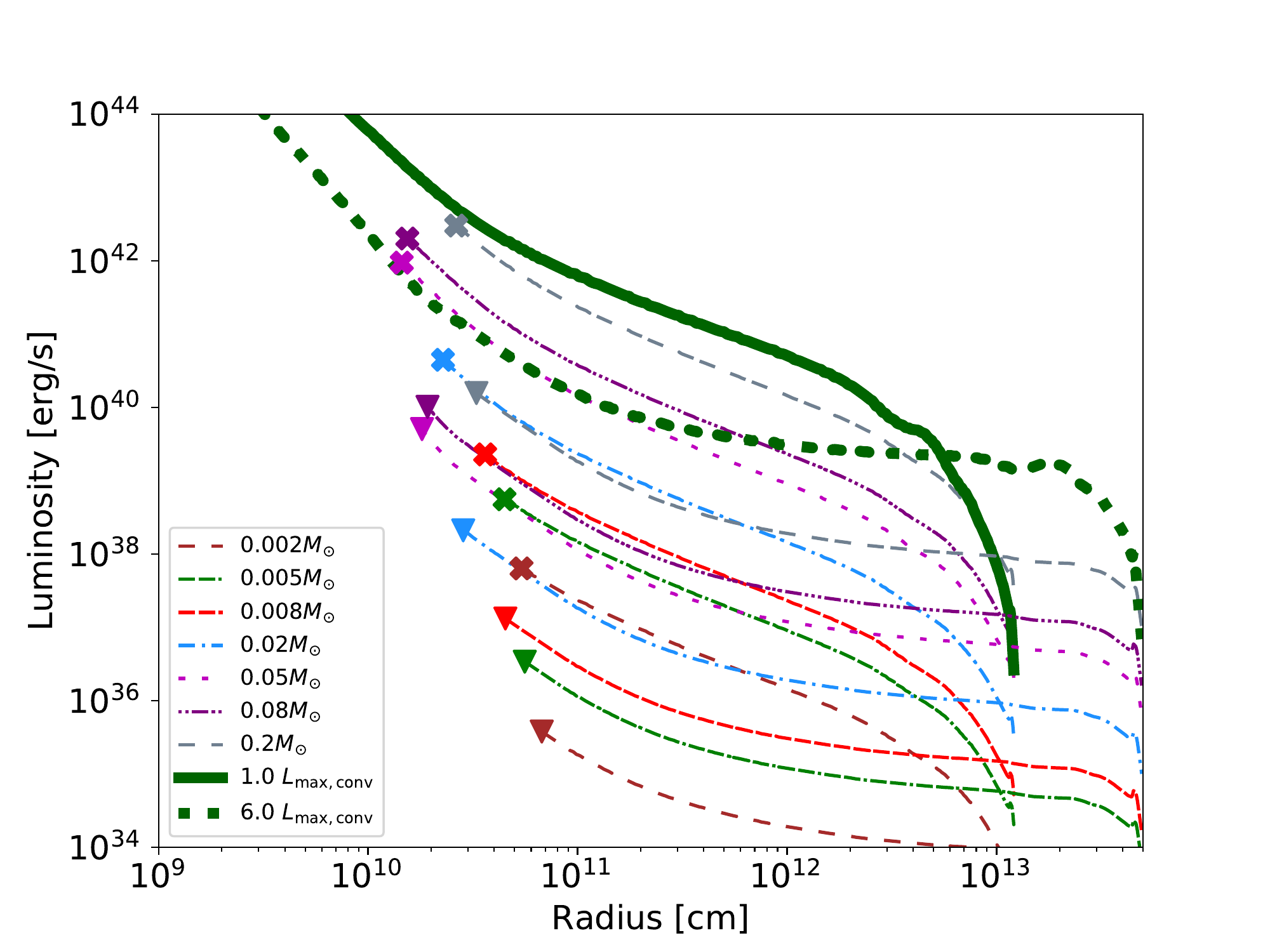}
   \caption{The maximum luminosity carried by convection, $L_{\rm max,conv}$ for two primary masses in thick green curves, is shown with the drag luminosity of several companions for the two primary mass cases in dashed, coloured lines. The tidal disruption radii are marked with an X for the $1.0 M_{\odot}$ case and a triangle for the $6.0 M_{\odot}$ case. This demonstrates that convection can carry the liberated orbital energy to the surface where it is radiated away.}
    \label{fig:lum}
\end{figure}

Within the primary's convective envelope, subsonic convection can accommodate additional power up to a maximum defined by:
\begin{equation}
     	L_{\rm max,conv} = \beta 4 \pi \rho[r] r^2 c_s^3[r],
	\label{Lconv}
\end{equation}
where $\beta \simeq 5$ and $\rho[r]$ and $c_s[r]$ are the density and the sound speed of the envelope medium at radius $r$, respectively \citep{Grichener2018a}. If the additional luminosity generated from inspiral remains below this maximum, convection can transport the energy to other regions of the star. The drag luminosity, generated from inspiral, is given as:
\begin{equation}
	L_{\rm drag} = \xi \pi r_{\rm acc}^2 \rho[r] v_{\phi}^3[r]
	\label{Ldrag}
\end{equation} 
        where the accretion radius is $r_{\rm acc} = 2 G m_2/v_{\phi}^2[r]$ \citep{Nordhaus2006}. 
        
        We compare the maximum luminosity carried by convection to the drag luminosity at each radius. For all mass ratios in this study, the drag luminosity is less than the limit that convection can carry, as shown in Figure~\ref{fig:lum}. The thick, solid and dotted green curves show the maximum luminosity that can be carried by the convective envelope for the $1.0 M_{\odot}$ and $6.0 M_{\odot}$ models, respectively. These two curves bound the convective luminosity limits for all primary masses in this work. The drag luminosities due to companions of mass $0.002 M_{\odot} - 0.2 M_{\odot}$ inspiraling through the two limiting primaries can be seen in the coloured, dashed curves. The tidal disruption radii are shown with X symbols and triangles for companions orbiting within the $1.0 M_{\odot}$ and $6.0 M_{\odot}$ primaries, respectively. 
        
        The solid green curve denoting the maximum luminosity carried by a $1.0 M_{\odot}$ primary exceeds all of the drag luminosity curves which are marked by an X, just as the dotted green curve denoting the maximum luminosity carried by a $6.0 M_{\odot}$ primary exceeds all of the drag luminosity curves which are marked by a triangle. Therefore, the luminosity produced during inspiral can be carried by convection for the primary-companion mass pairs in this study, as the two cases presented here are representative of the extrema cases.

\color{black}
\subsection{Common Envelope Outcomes}

There are two possible outcomes for common envelope phases: (i.) the companion survives the interaction and emerges in a short-period, post-CE binary, or (ii.) it does not and is destroyed in the process.

The companion body's radius, $R_2$, is estimated according to its mass.  For planet-mass objects ($m_2 \le 0.0026M_{\odot}$, \citealt{Zapolsky1969}) the radius is approximated as $R_2=R_{\mrm{Jupiter}}$. For brown dwarfs ($0.0026 M_{\odot} < m_2 < 0.077 M_{\odot}$, \citealt{Burrows1993}), the radius is calculated via:
\begin{equation}
    R_2/R_{\odot}=0.117-0.054\log^2\left({\frac{m_2}{0.0026 M_\odot}}\right)+0.024\log^3\left({\frac{m_2}{0.0026 M_\odot}}\right) \nonumber
    \label{eq:bdrad}
\end{equation} \citep{ReyesRuiz1999}.
For stellar main-sequence companions ($m_2 \ge 0.077M_{\odot}$), a power law is used:
\begin{equation}
R_2=\left(\frac{m_2}{M_{\odot}}\right)^{0.92} R_{\odot}
\label{eq:starrad}
\end{equation} \citep{ReyesRuiz1999}.

\begin{figure*}
\begin{multicols}{2}
 \centering
 \begin{minipage}{0.45\textwidth}
   \includegraphics[width=\textwidth]{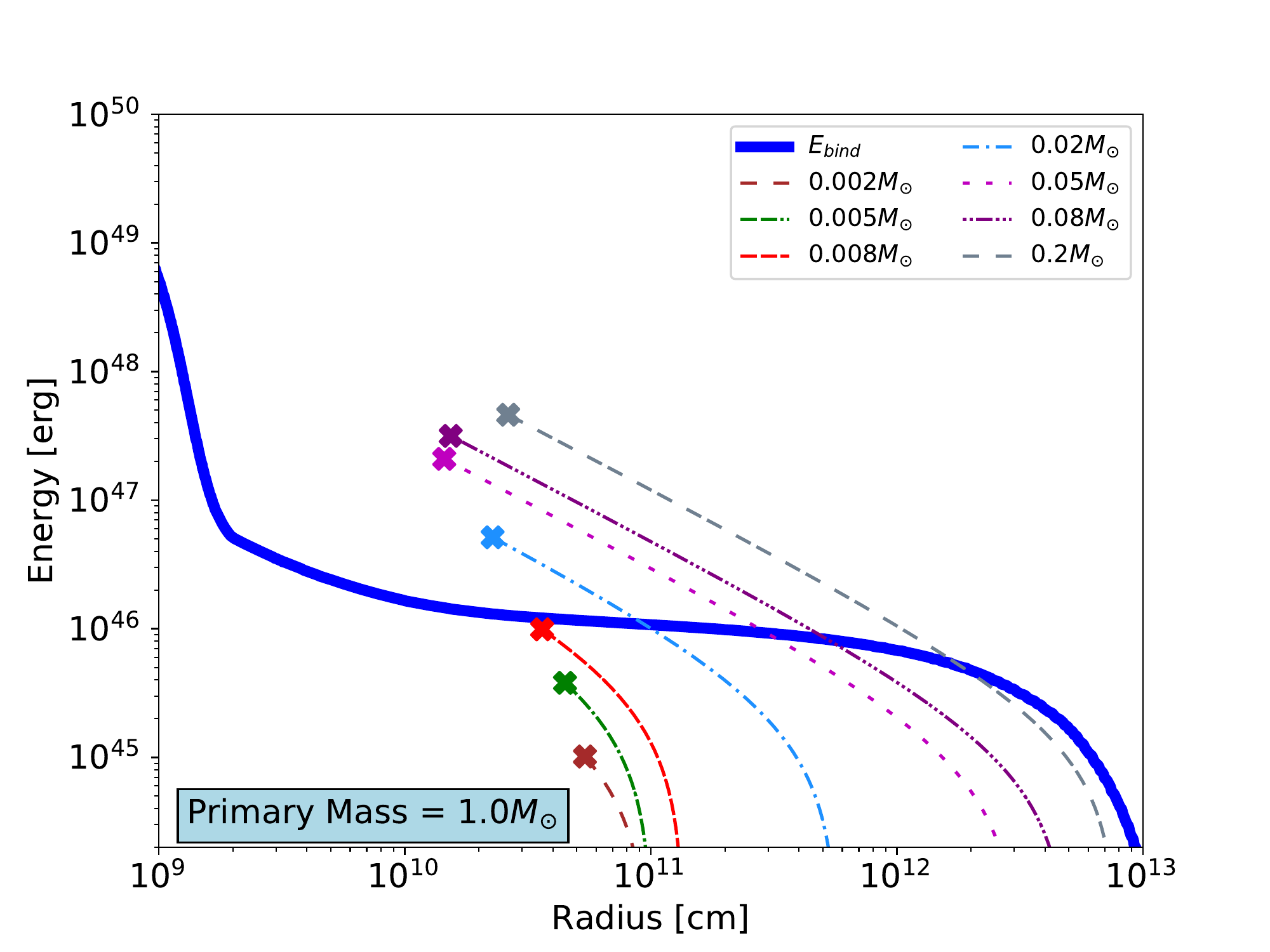}
 \end{minipage}
\hfill
 \begin{minipage}{0.45\textwidth}
   \includegraphics[width=\textwidth]{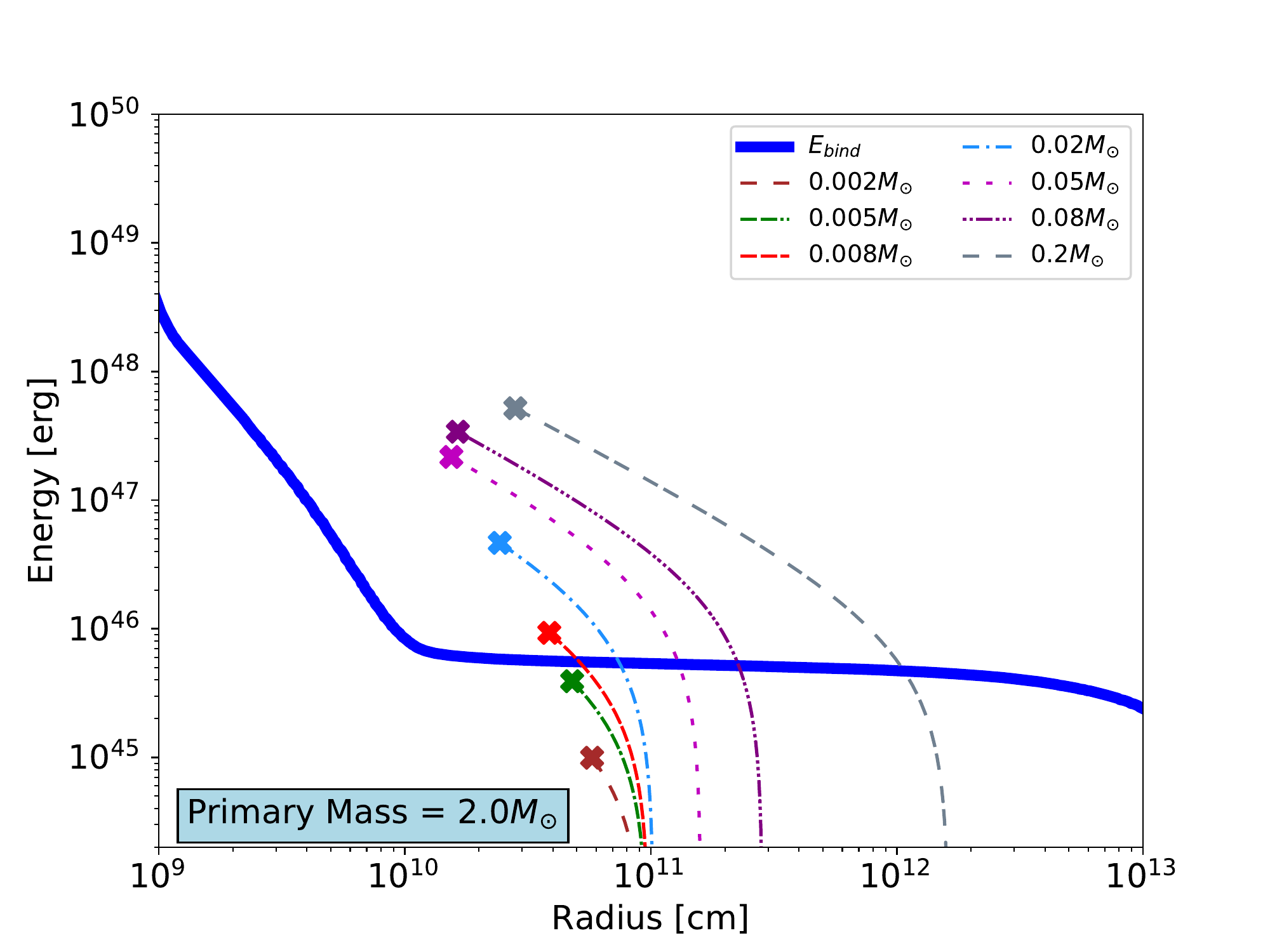}
 \end{minipage}
\hfill
 \begin{minipage}{0.45\textwidth}
   \includegraphics[width=\textwidth]{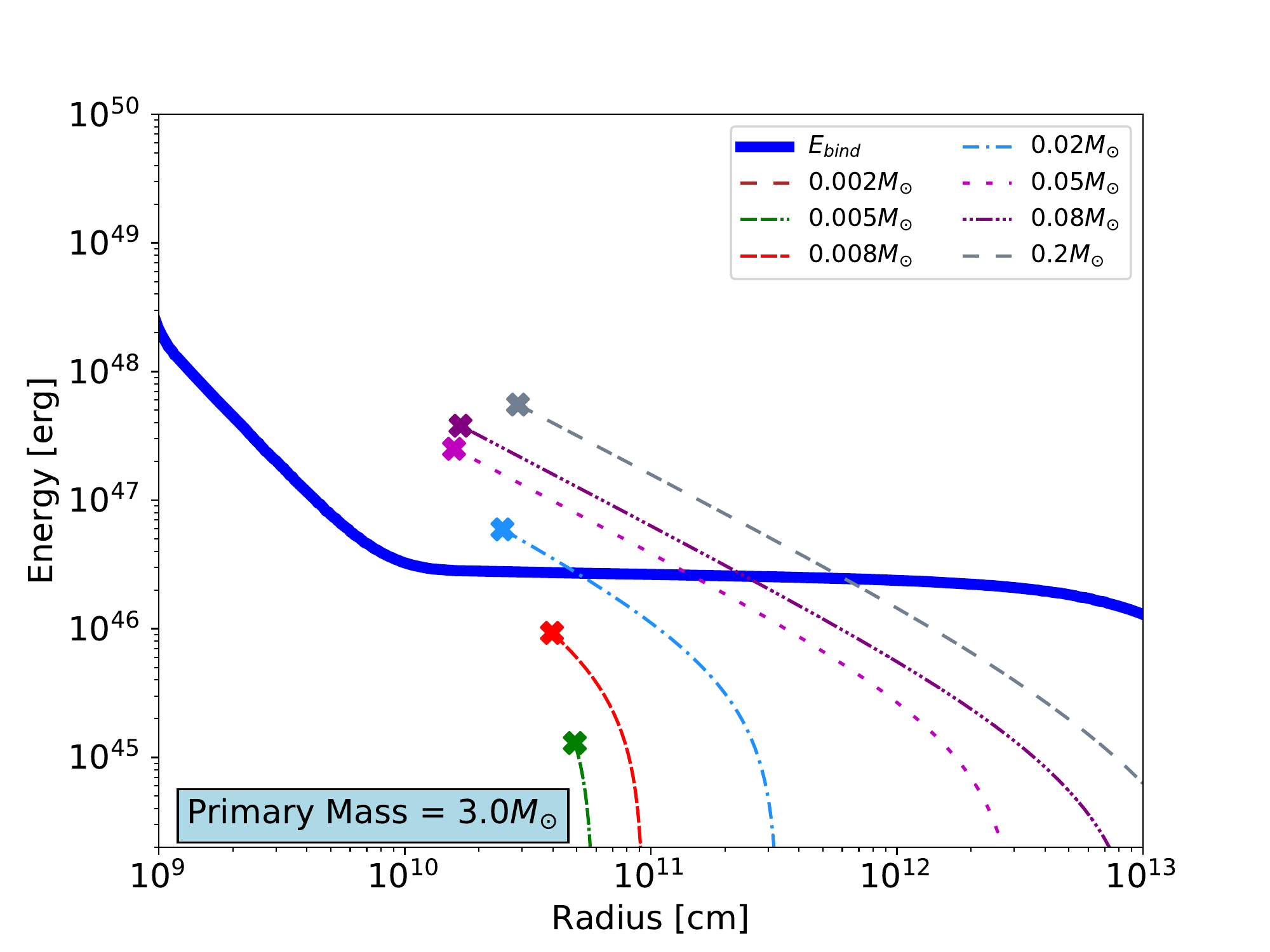}
 \end{minipage}
\hfill
 \begin{minipage}{0.45\textwidth}
   \includegraphics[width=\textwidth]{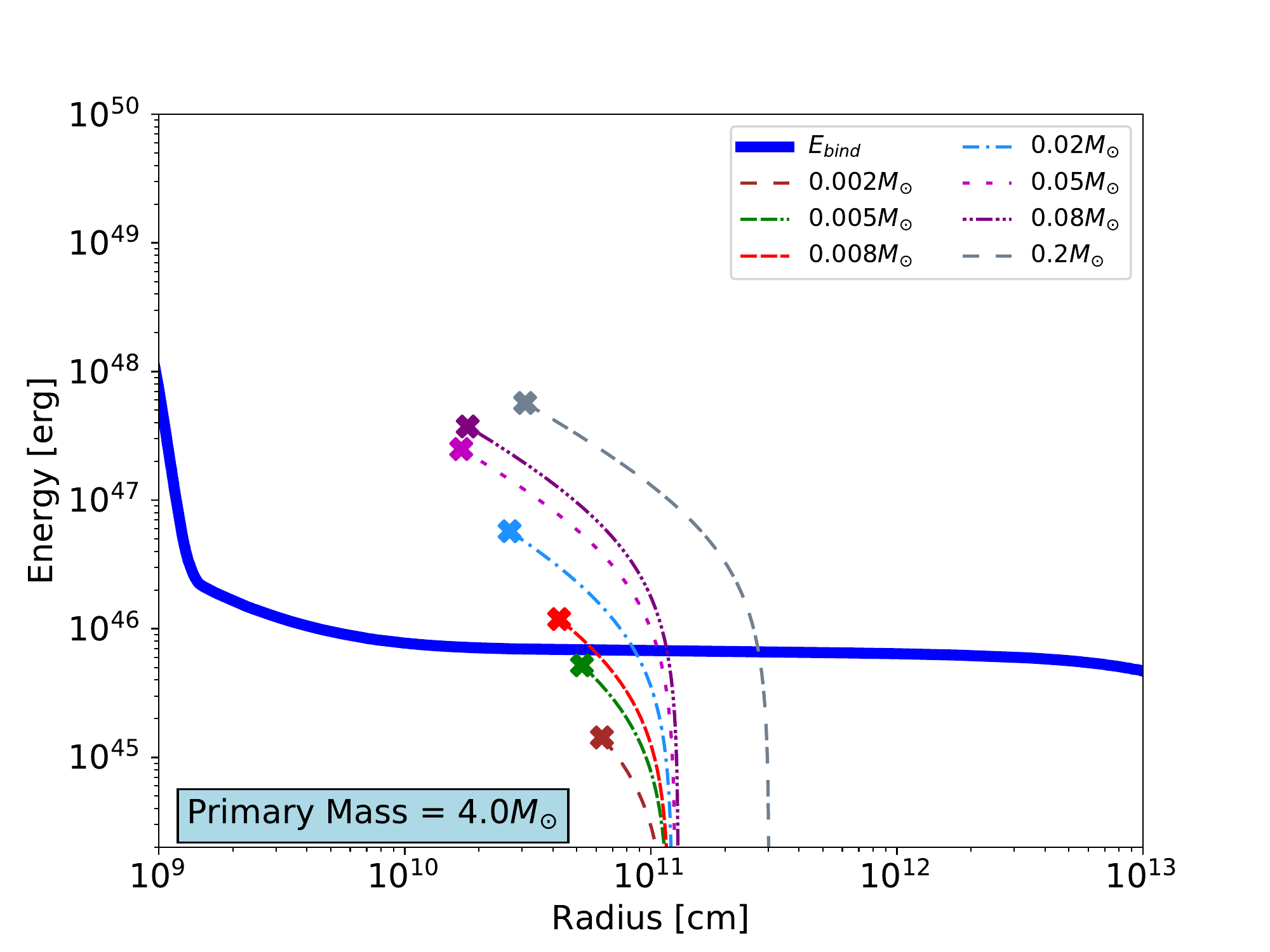}
 \end{minipage}
 \hfill
  \begin{minipage}{0.45\textwidth}
   \includegraphics[width=\textwidth]{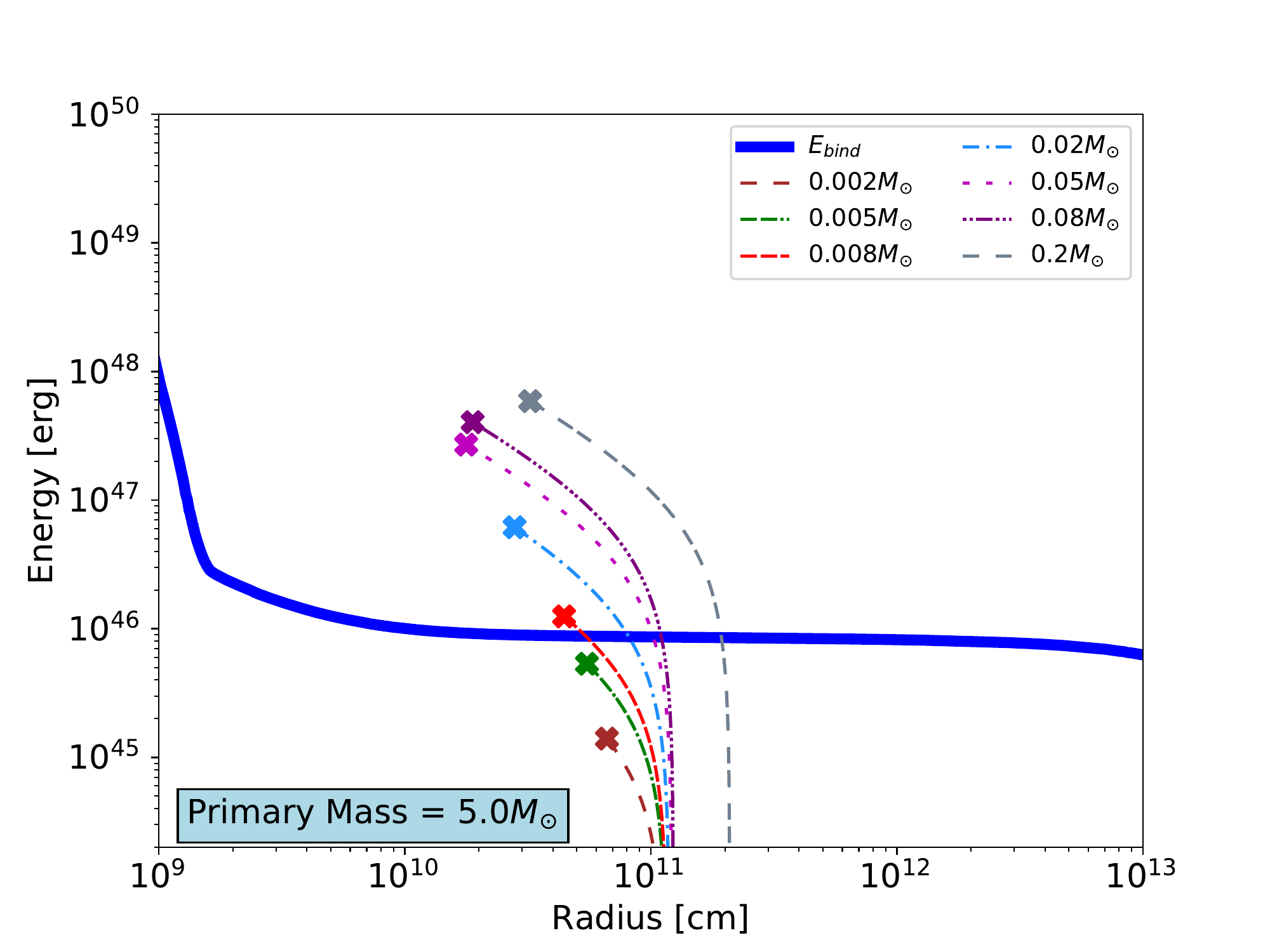}
 \end{minipage}
 \hfill
  \begin{minipage}{0.45\textwidth}
   \includegraphics[width=\textwidth]{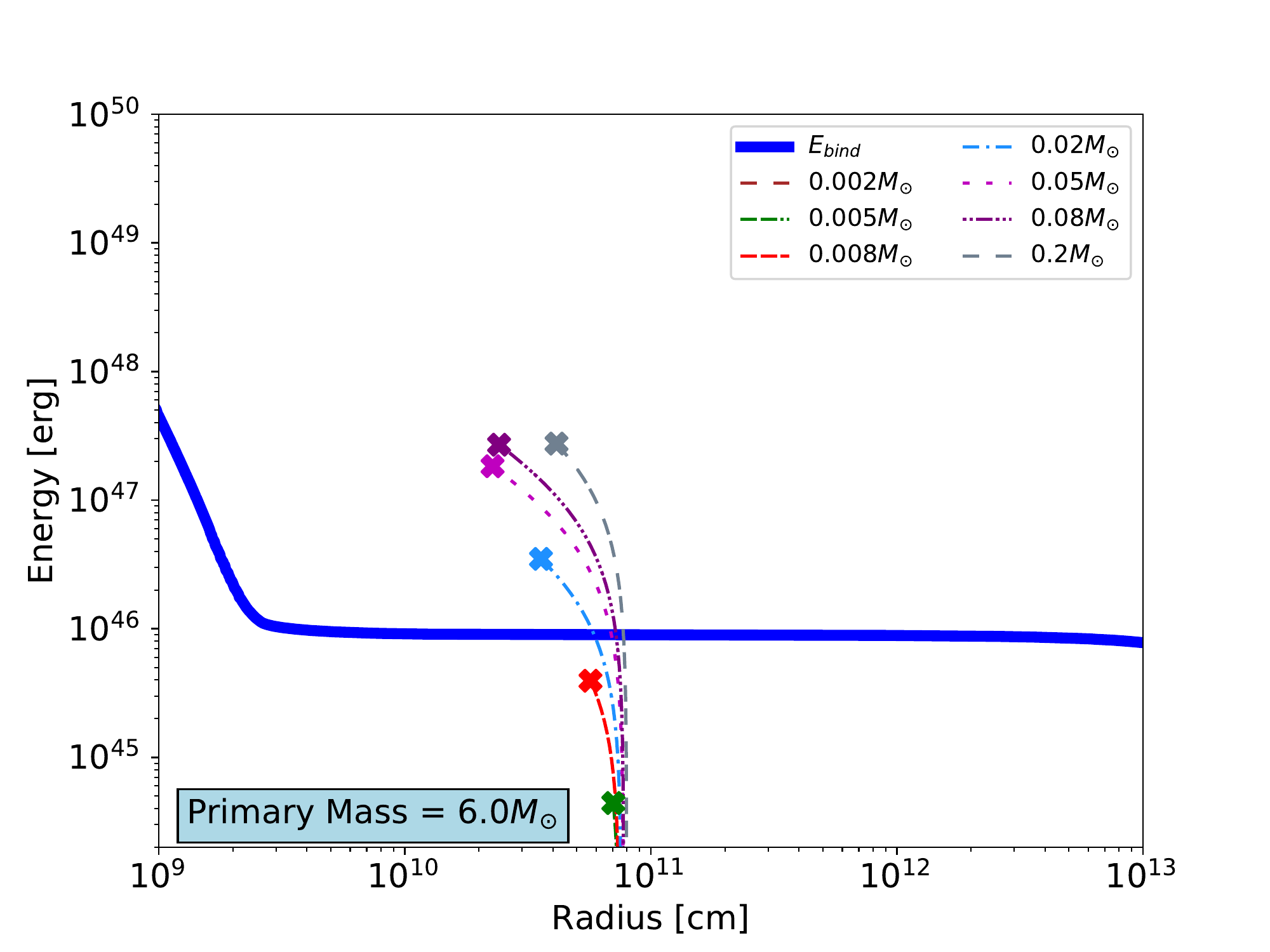}
 \end{minipage}
  \end{multicols}

 \hfill
 \caption{Comparative energy plots for a sample of representative primary masses at their maximum radial extent and several test companion masses. The binding energy for the primary star is shown in solid blue. The coloured, dashed lines show the change in orbital energy of the companion star as it inspirals, and the radius at which the companion shreds is marked with an X. (Several X's fall below $10^{44}$ erg.) For companion masses which the X falls below the binding energy curve, the companion will disrupt during inspiral before enough energy is transferred to unbind the envelope of the primary. These orbital energy curves take into account the convective zones of the primary, in that movement through the surface-contact convective regions (SCCRs) does not contribute energy to the ejecting of the envelope.
}
 \label{fig:AlphaEorbs}
\end{figure*}

To determine whether a companion survives the CE and emerges in a short-period orbit, there must exist an orbital separation, $a$, such that: (i.) $\abareff \Delta E_{\mathrm{orb}}[a] > E_{\mathrm{bind}}[a]$ and (ii.) $a > r_{\rm shred}$.  The shredding radius of each companion mass is determined as described in Section~\ref{sec:convregions}. The shredding radii are presented as X symbols in Figures~\ref{fig:Timescales} and~\ref{fig:AlphaEorbs}.  If the companion disrupts early in its descent through the envelope, the energy available to unbind the envelope will be minimized, whereas if the companion body remains intact through the majority of the envelope, the opposite will be true. This is discussed in more detail in Section~\ref{sec:transfer}.

\section{Ejection Efficiency in Convective Regions}
\label{sec:transfer}
The efficiency with which orbital energy can be used to unbind the envelope is a function of position inside the CE.  To determine a lower limit for $\abareff$ in a star with convective regions, we proceed in the following manner.  If $t_{\rm conv} < t_{\rm inspiral}$, and the companion is orbiting inside a surface-contact convective region (SCCR), then there is no contribution to $\abareff$ as the orbital energy can be transported via convection to the surface and leave the system as photons.  If the companion is orbiting in a region where $t_{\rm inspiral} < t_{\rm conv}$, then the orbital energy cannot escape the system and fully contributes, in some form, toward raising the negative binding energy of the primary.  In this way, we construct an average ejection efficiency by determining individual binary $\aeff$ coefficients (0 or 1) at each position.  We designate regions within the SCCR to have $\aeff=0$, as the energy can be carried to an optically thin layer and radiated away; we designate all other regions to have $\aeff=1$, as we assume that the energy transferred by the change in orbital energy of the companion gets evenly distributed throughout the mass in each radial shell. Then for each primary-companion pair, we determine $\abareff$ by integrating from the surface to either the point of tidal disruption or the point of energy equivalence (i.e. where $E_{\mathrm{bind}} = \aeff[r] \Delta E_{\mathrm{orb}}$),
 via:

\begin{equation}
\abareff = \frac{\int_{r_{\mrm{i}}}^{r_{\mrm{f}}}\aeff[r] \mrm{d}E_{\mrm{orb}}[r]}{E_{\mrm{orb}}[r_{\mrm{f}}]-E_{\mrm{orb}}[r_{\mrm{i}}]},
\label{eq:abar}
\end{equation}
where $\mrm{d}E_{\mrm{orb}}[r]$ can be calculated discretely as in Equation~\ref{eq:Eorb} and $r_{\mrm{f}}$ is the final position of the companion. The $r_{\mrm{f}}$ limit will be the maximum of $r_{\mrm{shred}}$ and $r_{E_{\mathrm{bind}} = \aeff[r] \Delta E_{\mathrm{orb}}}$. This limit is set due to the ejection of the envelope (when $r_{\rm f} = r_{E_{\mathrm{bind}} = \aeff[r] \Delta E_{\mathrm{orb}}}$) or due to the tidal disruption of the companion (when $r_{\rm f} = r_{\rm shred}$). The maximum of these two values is taken as the integral's upper limit, as the inspiral advances from larger radii towards the core.

Note that we assume the internal structure of the primary is constant during inspiral.  Since orbital energy is a function of enclosed mass of the primary and position, liberated orbital energy is distributed to the mass present in each location.

\begin{figure*}
  \includegraphics[width=1.0\textwidth]{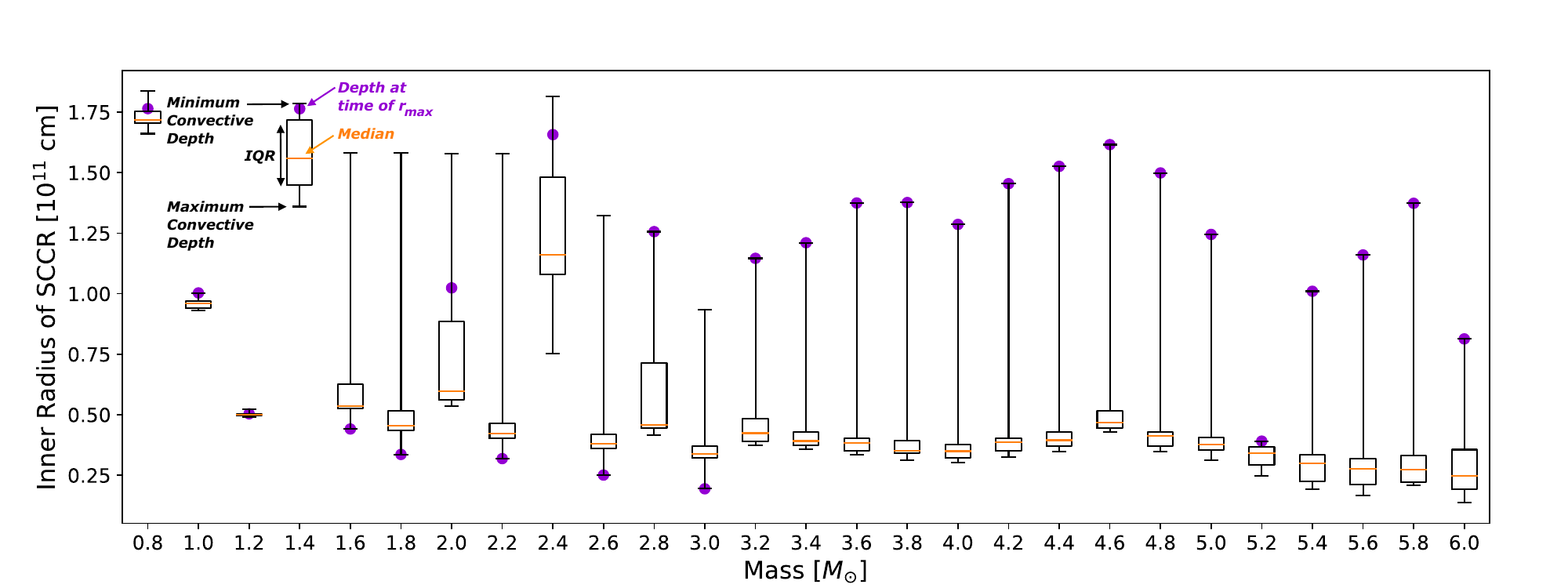}
    \caption{Box-and-whisker plot shows the range of convective depths during the final thermal pulse of the primary mass just prior to and including the time of maximum radius (does not include any time after the maximum radius). The orange line inside the box shows the median value of the boxplot, the vertical extent of the box marks the interquartile range (IQR: the middle 50\% of the range), and the upper and lower bounds of the whiskers mark the minimum and maximum convective depths, respectively. Note that for primary masses below $1.4 M_{\odot}$ the spanned range is very small, showing a stable convective region.
}
    \label{fig:boxplot}
\end{figure*}

The curves showing $E_{\mrm{bind}}$ and $\Delta E_{\mrm{orb}}$ for primaries between $1.0~{-}~6.0 M_{\odot}$ and companions between $0.002-0.2 M_{\odot}$ can be seen in Figure~\ref{fig:AlphaEorbs}. Each subplot shows the binding energy of the primary (thick blue line), compared with the change in orbital energy of several companions, in dashed, coloured lines. The transfer of the released orbital energy is halted by the SCCR ($\aeff=0$), resulting in low, unchanging values of $E_{\mrm{orb}}$, and resumes once the companion has inspiraled deeper than the lower boundary of the SCCR ($\aeff=1$). The radius at which the companion tidally shreds is marked with an X symbol, halting energy transfer.

On the six subplots in Figure~\ref{fig:AlphaEorbs}, the location of the X-symbols show which companion masses unbind the envelope; if the X falls above the solid blue $E_{\mrm{bind}}$ curve, the companion survives the inspiral and emerges as a post-CE binary in a short-period orbit. We see that for a representative SCCR depth of $10^{11}$ cm, companions between $0.008-0.02 M_{\odot}$ (${\sim}8-20 M_\mrm{Jupiter}$) and greater will successfully unbind the envelope and survive the binary interaction.

\subsection{Variability of the SCCR}
\label{sec:boxplot}

As described in Section~\ref{sec:convregions}, we argue that mixing of the energy released by the inspiraling companion occurs within the convective regions of the primary star, with emphasis on the potential for the SCCR to carry energy that is then radiated away. Therefore, an understanding of the variability of the SCCR is imperative to a complete understanding of patterns in the ejection efficiency. An examination of the stability of the SCCR with mass is shown in Figure~\ref{fig:boxplot}. This box-and-whisker plot displays the range of convective depths of the SCCR during the final thermal pulse of all monitored primary masses (boxes and whiskers), the median SCCR depth (orange line), and the SCCR depth at the time of maximum radius (purple circle). 

The variability of the SCCR becomes evident first in the $1.4M_{\odot}$ model and the range of SCCR depths remains large for all greater primary masses. The primary's time of maximum radius consistently corresponds with the minimum convective depth following and including the $3.2 M_{\odot}$ model, thus maximizing $\abareff$. For the instances where the SCCR is at maximum depth and maximum radius concurrently, the companion tidally shreds within the deep SCCR or shortly thereafter, minimizing $\abareff$.

\begin{figure}
	\includegraphics[width=\columnwidth]{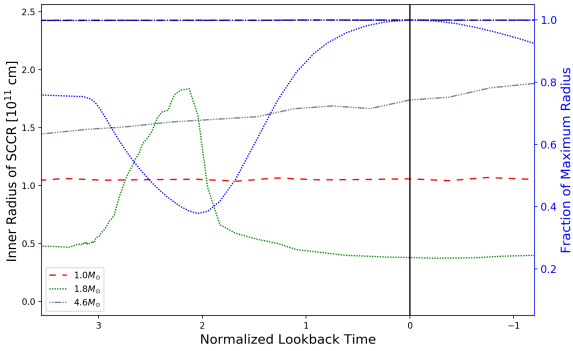}
    \caption{The SCCR depths over time are shown along with the primary's radius for three representative primary masses. The x-axis is a lookback time until maximum radius, described by: $\frac{t[r_{\mrm{max}}]-t[r]}{t[r_{\mrm{max}}]}$. The vertical black line marks the time of $r_{\mrm{max}}$, $t=0$. (Times after maximum radius are shown here for completeness but are not examined in this work.) The blue lines show the fraction of maximum radius in time. Two of the three blue lines overlap at unity. The coloured, dashed lines show the SCCR depth over time. Note the instability in the convective zone as it approaches the maximum radius for the $1.8 M_{\odot}$ model. (Convective depth is maximized with decreased interior convective radius.)
}
    \label{fig:convandradius}
\end{figure}

The SCCR depth over time is of interest because of its inconsistency at the time of maximum radius, which is evident in Figure~\ref{fig:boxplot}. The SCCR depth of three representative models are plotted over time and can be seen in Figure~\ref{fig:convandradius}. The $1.0M_{\odot}$ model, $1.8M_{\odot}$ model, and $4.6M_{\odot}$ model are representative of a stable SCCR, a maximum-SCCR-depth-at-$\mrm{r_{max}}$ SCCR, and a \textit{minimum}-SCCR-depth-at-$\mrm{r_{max}}$ SCCR, respectively.  The depth of the SCCR over time is plotted in the coloured, dashed lines. The respective radii of the primary over time are shown in blue, with dash patterns identical to those for the companion-mass SCCR depths. The lookback time is normalized to the maximum age of the star, and centred around the time of $\mrm{r_{max}}$, $t=0$. Each tick of normalized lookback time corresponds to a duration on the order of ${\sim}10^2$ years. For the $1.8M_{\odot}$ model, the time of peak radius corresponds with the time of deepest SCCR, thus minimizing the $\abareff$ value. The $1.0M_{\odot}$ SCCR depth remains remarkably constant, and the depth of the $4.6M_{\odot}$ SCCR decreases with time, in stark contrast to that of the $1.8 M_{\odot}$ model. Companions will have been engulfed by $\mrm{r_{max}}$ or will never be engulfed. Therefore, though the SCCR depth of the $4.6M_{\odot}$ model continues to decrease during the time after $\mrm{r_{max}}$ (see Fig.~\ref{fig:convandradius}), the SCCR is considered at ``minimum depth'' in Figure~\ref{fig:boxplot} as the interior structure of the primary is only of interest prior to and including the time at maximum radial extent.

\subsection{Ejection Efficiency, \boldmath{$\abareff$}}
\label{sec:ejectioneff}

\begin{figure}
\centering
\includegraphics[width=0.45\textwidth]{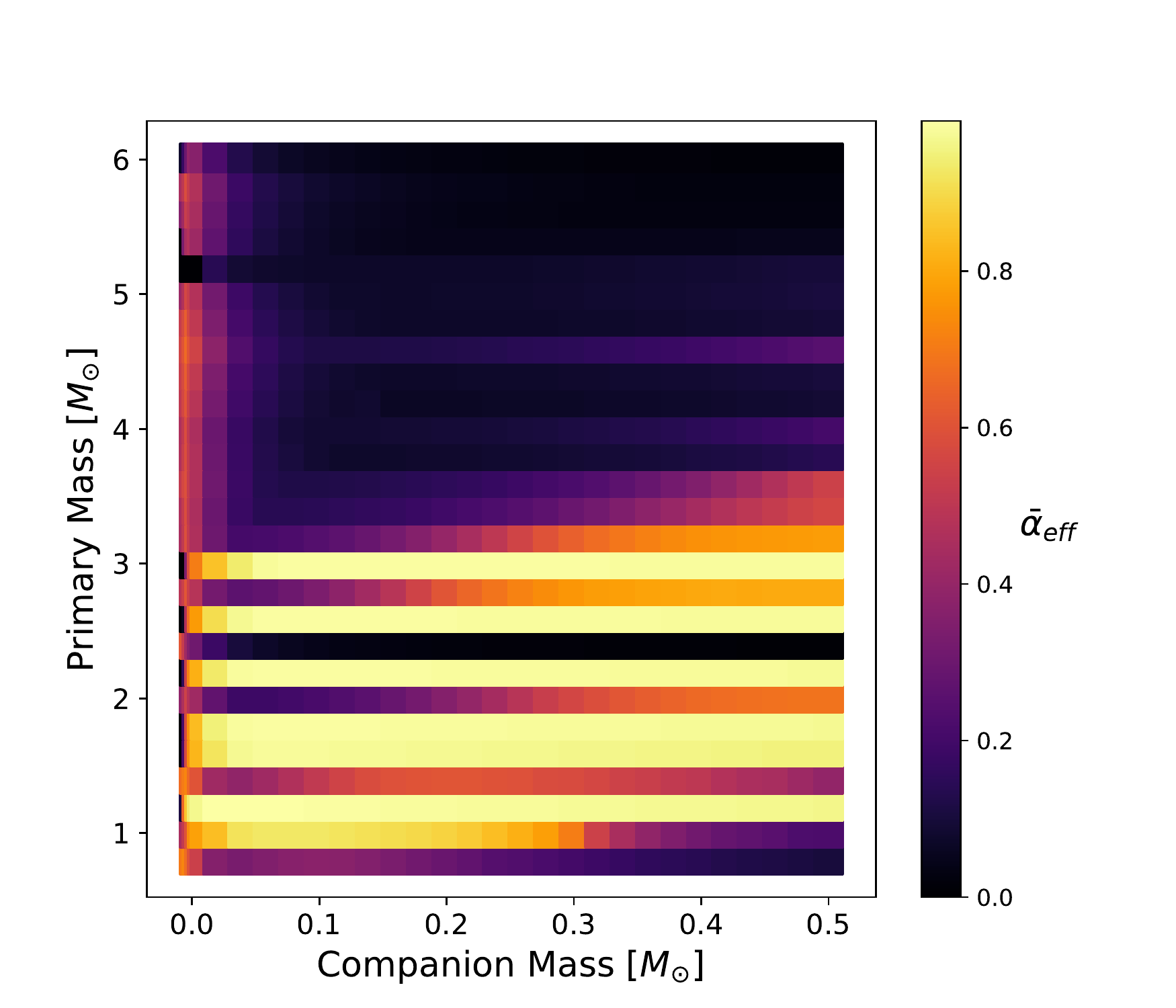}
   \caption{Colourmap of effective ejection efficiencies ($\abareff$) based on surface-contact convective regions (SCCRs) of the primary star.}
    \label{fig:colourmap}
\end{figure}

The ejection efficiency is unique for each primary-companion mass pair, since $\abareff$ depends on the specific internal stellar structure (especially the properties of the SCCR) and the properties of the companion.  We calculate $\abareff$ values for a matrix of primary-companion pairs (see Equation~\ref{eq:abar}) and present the results in Figure~\ref{fig:colourmap}.

The distinct horizontal stripes between primary masses of $1.2M_{\odot}$ and $3.0M_{\odot}$ are of interest. This phenomenon can be attributed to variability of the SCCR during the final thermal pulse (${\sim}10^2$ years) of the primary star thus making the ejection efficiency sensitive to the time of the CE phase. In cases where $\abareff{\sim}1$, 
the SCCR is relatively deep with long convective timescales (see, e.g., $3.0 M_{\odot}$ panel of Figure~\ref{fig:Timescales}). Since the inspiral timescale is shorter than the convective timescale, energy is very efficiently distributed throughout the envelope and maximizes $\abareff$. These cases also have a maximum SCCR depth at the time of maximum radius, unlike all other primary masses.

Low $\abareff$ values are seen in the upper half of the colourmap, where primary masses $>3 M_{\odot}$. In these cases, the lower edge of each SCCR is at approximately $10^{11}$ cm from the center of the primary, and the inspiral timescales are greater with increasing primary mass. Because of these two factors, energy can be carried more readily by convective transport and lower $\abareff$ in each case. A smoother spread of $\abareff$ would be expected for lower masses as well if the size of the SCCR were stable at the time of maximum radius.

The distinct features in the colourmap show the ejection efficiency's strong dependence on SCCR depth, which varies during post-main-sequence evolution (see Figures~\ref{fig:boxplot}~and~\ref{fig:convandradius}). Therefore, the ejection efficiency of these systems is sensitive to the time of interaction, or the age of the primary when the CE phase occurs.

\section{Discussion of Results} 
\label{sec:massratio}

\begin{figure}
	\begin{minipage}{\textwidth}
	\includegraphics[width=0.55\columnwidth]{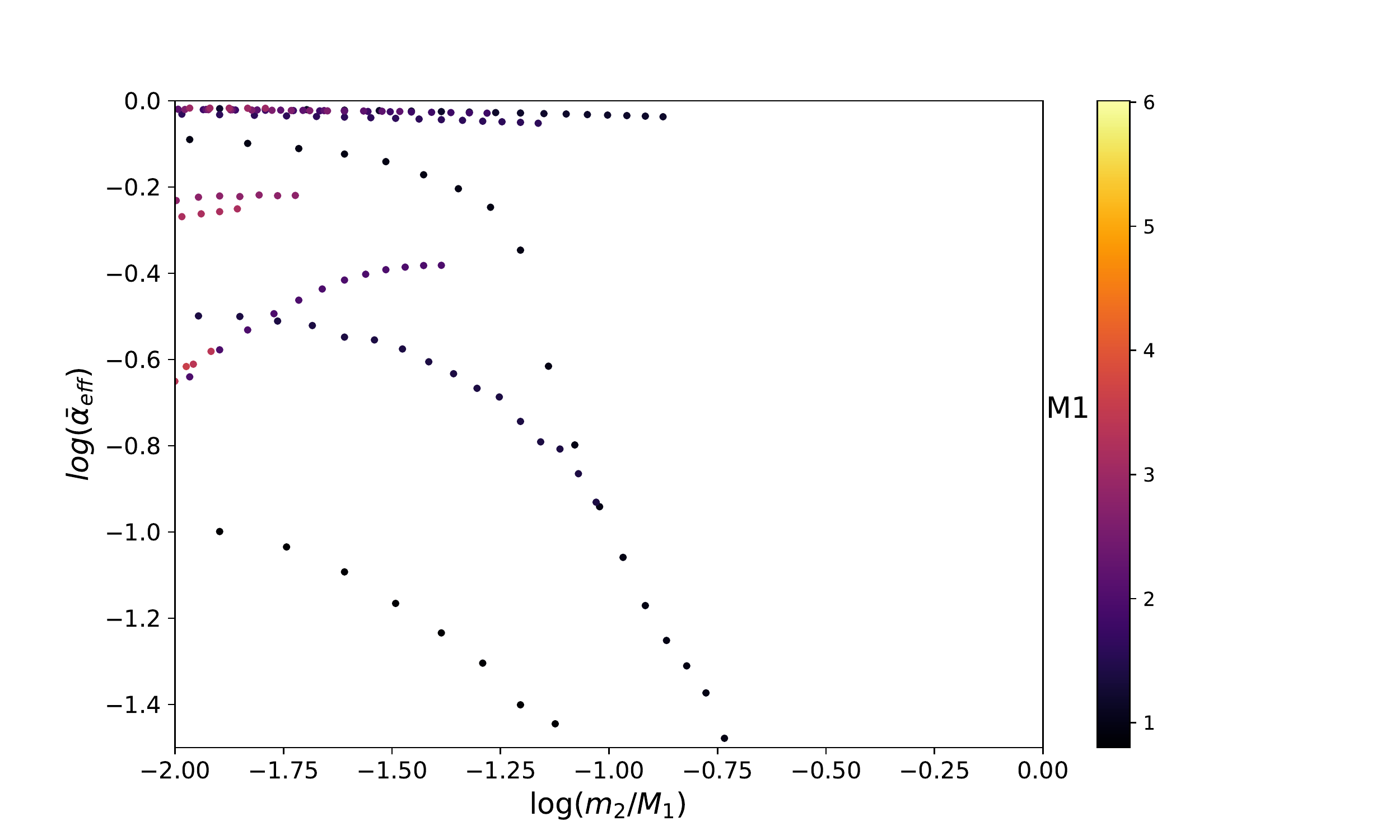}
	\end{minipage}
	\hfill
	\begin{minipage}{\textwidth}
	\includegraphics[width=0.55\columnwidth]{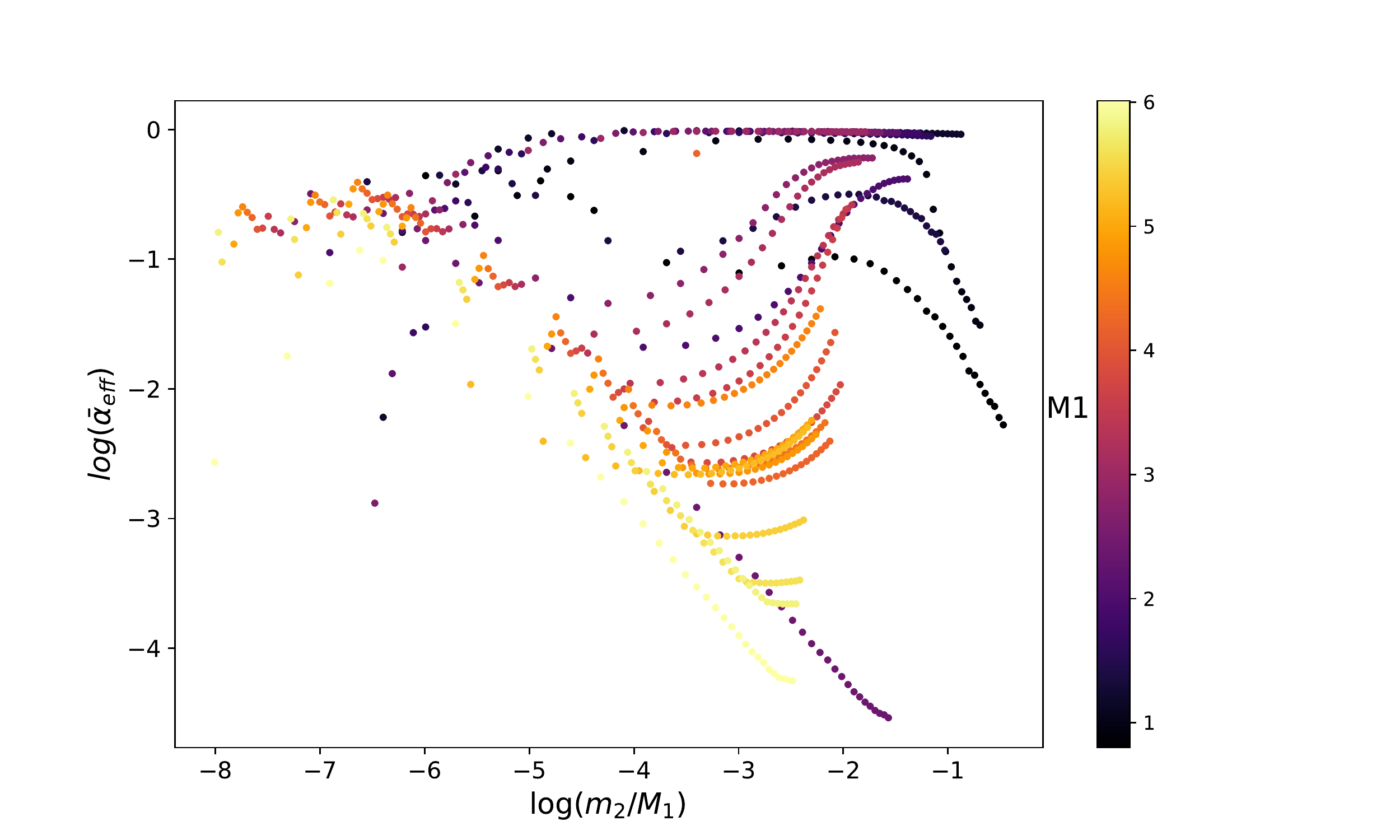}
	\end{minipage}
	\hfill
    \caption{Top: Axis-constrained natural logs of mass ratio and corresponding $\abareff$. These axis limits are comparable to the parameter space examined by \citet{DeMarco2011}, who also found a negative slope in this range.  Bottom: Mass ratio and $\abareff$ for all primary-companion mass pairs in this study. }
    \label{fig:mass_ratio}
\end{figure}

In comparison to observational results presented by \citet{DeMarco2011} and \citet{Zorotovic2011}, we plot the $\abareff$ vs. $q=m_2/M_1$ relation found from our simulated data in Figure~\ref{fig:mass_ratio}.  The top panel is limited to the parameter space considered by \citeauthor{DeMarco2011} while the parameter space for our full suite of primary-companion mass pairs are shown in the bottom panel. Within the range of \citeauthor{DeMarco2011}, many of our ejection efficiencies do show an anti-correlation with mass ratio but with varying slopes.  \citet{Zorotovic2011} are, however, in disagreement with \citeauthor{DeMarco2011}, finding larger final separations with lower companion masses. In the lower panel of Figure~\ref{fig:mass_ratio}, in which we present our full results, there are regions of parameter space that positively correlate with mass ratio.

It is worthwhile to note that \citeauthor{DeMarco2011} and \citeauthor{Zorotovic2011} estimate $\abareff$ through observations of assumed CE progenitors, whereas we use stellar evolution models to probe the interior structure of primaries during the CE phase. For this reason, we note that a direct comparison is difficult as both studies estimate the CE binding energy via a $\lambda$ parameter. This can result in ejection efficiencies that are greater than unity making direct comparison challenging.

\citet{Politano2007} argue that $\abareff$ is a function of the companion mass and the interior structure of the evolved primary, a statement with which we agree. Through this work, we find that the ejection efficiency is, in fact, highly sensitive to properties of the convective regions of the primary during the CE phase. To simulate CE systems and find the $\abareff$ value, one must consider the effect of mixing within convective regions.
% Inspiral timescale also dependent on physical structure of envelope.

Convective mixing can affect the system in different ways depending on where it occurs. As the companion inspirals, convection can transport the released orbital energy away from the companion and distribute it throughout other regions. If convection occurs in the SCCR, then the convective eddies can carry the energy to the surface where it can be radiated away. This work assumes that all orbital energy released within the SCCR is radiated away and thus provides a lower limit on $\abareff$ under the assumption that the liberated orbital energy is distributed evenly among the mass in a given layer.

In some cases, the $\abareff$ value of a specific system may deviate a bit from the values presented here, due to internal structure changes during inspiral (the depth of the SCCR can vary on ${\sim}10^2$ year timescales, as in Figure~\ref{fig:convandradius}, and the estimated duration of the entire CE phase is of comparable length). For this work, we assumed the internal structure to remain constant once the companion was engulfed. In any case, $\abareff$ is sensitive to the time of the companion's inspiral through the envelope of primary, since the SCCR varies so rapidly during the evolution of a star. For this work, we used the profile of a primary at its maximum radius, and assumed that the secondary began to skim the primary's surface at that time. We chose this because at the maximum extent of the primary, the largest spatial volume in which tidal dissipation may lead to CE phases is the greatest.   Given our findings of the dependence of the quickly-changing SCCR, the ejection efficiency values shown in Figure~\ref{fig:colourmap} cannot be generalized to those companion-mass pairs. Instead, the SCCR of the primary at the specific time of companion's engulfment must be known to calculate ejection efficiencies for unique configurations.

\subsection{Implications of Convection}

Convection allows the binary to naturally shrink to short orbital periods before the liberated orbital energy can be tapped to drive ejection.  In many cases, these short periods are less than a day. This is, at least at initial glance, consistent with the steep drop off in observed post-CE systems that have orbital periods greater than a day \citep{Davis2010}.  Investigating the predicted post-CE population distributions when $\abareff$ from Equation~\ref{eq:abar} is adopted is an interesting future direction but requires determining when common envelope evolution starts. 

The final state of the system also depends largely on the dominant timescale for the companions. If the inspiral timescale is shorter than the convective timescale, like those seen in $1.0-3.0 M_{\odot}$ panels of Figure~\ref{fig:Timescales}, the final orbital separations of the more massive companion bodies result in ${\sim}3$-day periods and the final separations of the less massive companions that exceed the energy required to unbind the envelope result in $\lesssim 1$-day periods. The intersection of the energy curves in Figure~\ref{fig:AlphaEorbs} shows where the envelope will be ejected, and thus the final orbital separation.

In some numerical simulations, the rate of orbital decay is slowed due to the gas reaching co-rotation \citep{Ricker2012,Ohlmann2015,Chamandy:2018aa}.  However, co-rotation cannot be perfectly maintained in a turbulent medium and thus may lead to faster decay than is currently seen in such simulations. If the orbital decay timescale remains above the convective timescale in the SCCR, then the effect will be minimal as energy transport to the surface is the dominant process.  

Note that there are several effects which we have neglected that may significantly increase the orbital decay timescale, and thus further increase the importance of convective effects.  During inspiral, we have assumed that the gas is stationary and thus does not spin-up and reach near co-rotation with the orbit as is seen in some numerical simulations.  If the gas is indeed near co-rotation with the orbit, then the orbital decay timescale is significantly larger than the values presented in this work.  If that is the case, then even for our lowest-mass primaries, convection may dominate and carry the liberated orbital energy to the surface where it can be radiated away.  In a similar vein, we have also assumed that the inspiral has no effect on the convection itself.  However, the transfer of orbital energy to the gas during inspiral may result in larger convective velocities.  Such an effect would shorten the convection transport timescales.  The results presented here are conservative and would be improved by including both effects in future studies.

In regions where the orbital decay timescale is shorter than the convective turnover times, $\aeff[r]=1$.  In principle, this means that we are implicitly assuming that the orbital energy is equally distributed to the mass in that region.  At the inner boundary of the SCCR where this condition is satisfied, turbulent mixing may distribute sufficient orbital energy to enough of the mass to eject the full envelope.  Primaries with masses $>3.0 M_\odot$ have secondary convective zones that may also aid in mixing sufficient orbital energy for those companions that reach it before being disrupted.  These assumptions warrant investigation via numerical simulations that include convection (see \citealt{Chamandy:2018ab} for further discussion).

\section{Conclusions and future directions}

We have studied the effects of convection on the ejection efficiencies of common envelope interactions.  Using detailed stellar evolution models at the time of maximal radial extent, we calculate $\abareff$ values for a matrix of primary-companion mass pairs.  The ejection efficiencies are most sensitive to the properties of the surface-contact convective region (SCCR).  In this region, the orbital decay timescales are longer than the convective timescales, thereby allowing the star to effectively radiate released orbital energy and thus lower $\abareff$.  The inclusion of convection in CEs may solve the ejection problem seen in numerical simulations without the need for additional energy sources as the orbit must decay substantially before orbital energy can be tapped to drive ejection.  Our considerations of convection also allow for post-CE orbital periods of less than a day in higher primary masses, an observational result that has been infrequently reproduced in population synthesis models that use universal, or constant, ejection efficiencies.  

The results described are conservative, as changes in the envelope's gas are not considered during inspiral. For this reason, the co-rotation seen in numerical simulations cannot play a role in increasing the inspiral timescale, which may allow systems with even our lowest primary masses to end in sub-day orbital periods, just as our higher mass primary masses do. We also assume that the inspiraling companion's energy transfer does not affect the convective velocities. The released orbital energy may increase the convective velocities, consequently shortening the convective timescale. This, too, would strengthen the effect of convection on the system.

We provide a simple method to calculate $\abareff$ if the properties of the SCCR are known.  Since the ejection efficiencies are sensitive to the depth of the SCCR, they are inherently sensitive to the time of engulfment.  If the SCCR depth changes substantially over time for a given stellar evolution model, then the time of the CE onset will determine the ejection efficiency.  However, since RGB/AGB stars possess deep convective envelopes, the effects of convection remain important, independent of when a CE commences.

Future work can be advanced on multiple fronts.  A more comprehensive study of the effects of co-rotation on the inspiral timescale and thus the ejection efficiency should be carried out. Numerical work should include high-resolution simulations of convection in common envelopes.  Since this may be challenging in global simulations without convective sub-grid models, high-resolution local simulations of convective energy transport in stratified wind tunnels may be a natural starting point \citep{MacLeod:2017fk}.  Massive stars also host deep convective zones and the impact on the ejection efficiencies should be investigated in the context of formation channels for the progenitors of gravitational-wave driven, compact-object mergers \citep{Belczynski:2016sf}.  Finally, coupling these calculations to dynamical calculations that determine the time of engulfment can result in improved ejection efficiencies, which could then be incorporated into studies of populations \citep{Belczynski:2002nr,Moe:2006rm}.

\section*{Acknowledgements}
ECW and JN acknowledge support from the following grants: NASA HST-AR-15044, NASA HST-AR-14563 and NTID SPDI-15992.  The authors thank Gabriel Guidarelli, Jeff Cummings, Joel Kastner, Luke Chamandy, Eric Blackman, John Whelan and Adam Frank for stimulating discussions.

%%%%%%%%%%%%%%%%%%%%%%%%%%%%%%%%%%%%%%%%%%%%%%%%%%

%%%%%%%%%%%%%%%%%%%% REFERENCES %%%%%%%%%%%%%%%%%%

\bibliography{MNRASpaper} 

\begin{thebibliography}{}

\bibitem[\protect\astroncite{{Belczynski} et~al.}{2016}]{Belczynski:2016sf}
{Belczynski}, K., {Holz}, D.~E., {Bulik}, T., and {O'Shaughnessy}, R.: 2016,
\newblock {\em \nat} {\bf 534}, 512

\bibitem[\protect\astroncite{{Belczynski} et~al.}{2002}]{Belczynski:2002nr}
{Belczynski}, K., {Kalogera}, V., and {Bulik}, T.: 2002,
\newblock {\em \apj} {\bf 572}, 407

\bibitem[\protect\astroncite{{Bloecker}}{1995}]{Bloecker:1995aa}
{Bloecker}, T.: 1995,
\newblock {\em \aap} {\bf 297}, 727

\bibitem[\protect\astroncite{Burrows et~al.}{1993}]{Burrows1993}
Burrows, A., Hubbard, W.~B., Saumon, D., and Lunine, J.~I.: 1993,
\newblock {\em The Astrophysical Journal} {\bf 406}, 158

\bibitem[\protect\astroncite{{Canals} et~al.}{2018}]{Canals:2018aa}
{Canals}, P., {Torres}, S., and {Soker}, N.: 2018,
\newblock {\em \mnras} {\bf 480}, 4519

\bibitem[\protect\astroncite{{Chamandy} et~al.}{2018a}]{Chamandy:2018aa}
{Chamandy}, L., {Frank}, A., {Blackman}, E.~G., {Carroll-Nellenback}, J.,
  {Liu}, B., {Tu}, Y., {Nordhaus}, J., {Chen}, Z., and {Peng}, B.: 2018a,
\newblock {\em \mnras} {\bf 480}, 1898

\bibitem[\protect\astroncite{{Chamandy} et~al.}{2018b}]{Chamandy:2018ab}
{Chamandy}, L., {Yisheng}, T., {Blackman}, E.~G., {Carroll-Nellenback}, J.,
  {Frank}, A., {Liu}, B., and {Nordhaus}, J.: 2018b,
\newblock {\em in prep}

\bibitem[\protect\astroncite{{Chen} et~al.}{2017}]{Chen:2017aa}
{Chen}, Z., {Frank}, A., {Blackman}, E.~G., {Nordhaus}, J., and
  {Carroll-Nellenback}, J.: 2017,
\newblock {\em MNRAS} {\bf 468}, 4465

\bibitem[\protect\astroncite{Claeys et~al.}{2014}]{Claeys2014}
Claeys, J. S.~W., Pols, O.~R., Izzard, R.~G., Vink, J., and Verbunt, F. W.~M.:
  2014,
\newblock {\em A{\&}A} {\bf 563}, 83

\bibitem[\protect\astroncite{{Cojocaru} et~al.}{2017}]{Cojocaru:2017aa}
{Cojocaru}, R., {Rebassa-Mansergas}, A., {Torres}, S., and
  {Garc{\'{\i}}a-Berro}, E.: 2017,
\newblock {\em \mnras} {\bf 470}, 1442

\bibitem[\protect\astroncite{Davis et~al.}{2010}]{Davis2010}
Davis, P.~J., Kolb, U., and Willems, B.: 2010,
\newblock {\em Monthly Notices of the Royal Astronomical Society} {\bf 403(1)},
  179

\bibitem[\protect\astroncite{{De Marco} et~al.}{2011}]{DeMarco2011}
{De Marco}, O., Passy, J.-C., Moe, M., Herwig, F., Low, M.-M.~M., and Paxton,
  B.: 2011,
\newblock {\em Mon. Not. R. Astron. Soc} {\bf 411}, 2277

\bibitem[\protect\astroncite{{Fabrycky} and {Tremaine}}{2007}]{Fabrycky:2007aa}
{Fabrycky}, D. and {Tremaine}, S.: 2007,
\newblock {\em \apj} {\bf 669}, 1298

\bibitem[\protect\astroncite{{Glanz} and {Perets}}{2018}]{Glanz:2018aa}
{Glanz}, H. and {Perets}, H.~B.: 2018,
\newblock {\em \mnras} {\bf 478}, L12

\bibitem[\protect\astroncite{Grichener et~al.}{2018}]{Grichener2018a}
Grichener, A., Sabach, E., and Soker, N.: 2018,
\newblock {\em Monthly Notices of the Royal Astronomical Society} {\bf 478(2)},
  1818

\bibitem[\protect\astroncite{{Han} et~al.}{1995}]{Han:1995aa}
{Han}, Z., {Podsiadlowski}, P., and {Eggleton}, P.~P.: 1995,
\newblock {\em \mnras} {\bf 272}, 800

\bibitem[\protect\astroncite{Iben and Livio}{1993}]{Iben1993}
Iben, I. and Livio, M.: 1993,
\newblock {\em {December Common Envelopes in Binary Star Evolution 1}},
\newblock Technical report

\bibitem[\protect\astroncite{{Iben, I.} and Tutukov}{1984}]{Iben1984}
{Iben, I.}, J. and Tutukov, A.~V.: 1984,
\newblock {\em The Astrophysical Journal} {\bf 284}, 719

\bibitem[\protect\astroncite{{Ivanova}}{2018}]{Ivanova:2018aa}
{Ivanova}, N.: 2018,
\newblock {\em \apjl} {\bf 858}, L24

\bibitem[\protect\astroncite{{Ivanova} et~al.}{2013}]{Ivanova:2013aa}
{Ivanova}, N., {Justham}, S., {Avendano Nandez}, J.~L., and {Lombardi}, J.~C.:
  2013,
\newblock {\em Science} {\bf 339}, 433

\bibitem[\protect\astroncite{Ivanova et~al.}{2013}]{Ivanova2013}
Ivanova, N., Justham, S., Chen, X., {De Marco}, O., Fryer, C.~L., Gaburov, E.,
  Ge, H., Glebbeek, E., Han, Z., Li, X.~D., Lu, G., Marsh, T., Podsiadlowski,
  P., Potter, A., Soker, N., Taam, R., Tauris, T.~M., {Van Den Heuvel}, E.~P.,
  and Webbink, R.~F.: 2013,
\newblock {\em Astronomy and Astrophysics Review} 21(1)

\bibitem[\protect\astroncite{{Ivanova} et~al.}{2015}]{Ivanova:2015aa}
{Ivanova}, N., {Justham}, S., and {Podsiadlowski}, P.: 2015,
\newblock {\em \mnras} {\bf 447}, 2181

\bibitem[\protect\astroncite{{Kashi} and {Soker}}{2018}]{Kashi:2018aa}
{Kashi}, A. and {Soker}, N.: 2018,
\newblock {\em \mnras} {\bf 480}, 3195

\bibitem[\protect\astroncite{{Kochanek} et~al.}{2014}]{Kochanek:2014aa}
{Kochanek}, C.~S., {Adams}, S.~M., and {Belczynski}, K.: 2014,
\newblock {\em \mnras} {\bf 443}, 1319

\bibitem[\protect\astroncite{{Kruckow} et~al.}{2018}]{Kruckow:2018aa}
{Kruckow}, M.~U., {Tauris}, T.~M., {Langer}, N., {Kramer}, M., and {Izzard},
  R.~G.: 2018,
\newblock {\em \mnras} {\bf 481}, 1908

\bibitem[\protect\astroncite{{Kuruwita} et~al.}{2016}]{Kuruwita:2016aa}
{Kuruwita}, R.~L., {Staff}, J., and {De Marco}, O.: 2016,
\newblock {\em \mnras} {\bf 461}, 486

\bibitem[\protect\astroncite{Livio and Soker}{1988}]{Livio1988}
Livio, M. and Soker, N.: 1988,
\newblock {\em The Astrophysical Journal} {\bf 329}, 764

\bibitem[\protect\astroncite{{MacLeod} et~al.}{2017}]{MacLeod:2017fk}
{MacLeod}, M., {Antoni}, A., {Murguia-Berthier}, A., {Macias}, P., and
  {Ramirez-Ruiz}, E.: 2017,
\newblock {\em \apj} {\bf 838}, 56

\bibitem[\protect\astroncite{{MacLeod} et~al.}{2018}]{MacLeod:2018aa}
{MacLeod}, M., {Ostriker}, E.~C., and {Stone}, J.~M.: 2018,
\newblock {\em ApJ} {\bf 863}, 5

\bibitem[\protect\astroncite{Michaely and Perets}{2016}]{Michaely2016}
Michaely, E. and Perets, H.~B.: 2016,
\newblock {\em Monthly Notices of the Royal Astronomical Society} {\bf 458(4)},
  4188

\bibitem[\protect\astroncite{Michaely and Perets}{2018}]{Michaely2018}
Michaely, E. and Perets, H.~B.: 2018,
\newblock {\bf arxiv:1810.09454v1}, 1

\bibitem[\protect\astroncite{{Moe} and {De Marco}}{2006}]{Moe:2006rm}
{Moe}, M. and {De Marco}, O.: 2006,
\newblock {\em \apj} {\bf 650}, 916

\bibitem[\protect\astroncite{Nordhaus and Blackman}{2006}]{Nordhaus2006}
Nordhaus, J. and Blackman, E.~G.: 2006,
\newblock {\em Monthly Notices of the Royal Astronomical Society} {\bf 370(4)},
  2004

\bibitem[\protect\astroncite{{Nordhaus} et~al.}{2007}]{Nordhaus:2007aa}
{Nordhaus}, J., {Blackman}, E.~G., and {Frank}, A.: 2007,
\newblock {\em \mnras} {\bf 376}, 599

\bibitem[\protect\astroncite{{Nordhaus} and {Spiegel}}{2013}]{Nordhaus:2013aa}
{Nordhaus}, J. and {Spiegel}, D.~S.: 2013,
\newblock {\em \mnras} {\bf 432}, 500

\bibitem[\protect\astroncite{{Nordhaus} et~al.}{2010}]{Nordhaus:2010aa}
{Nordhaus}, J., {Spiegel}, D.~S., {Ibgui}, L., {Goodman}, J., and {Burrows},
  A.: 2010,
\newblock {\em MNRAS} {\bf 408}, 631

\bibitem[\protect\astroncite{{Nordhaus} et~al.}{2011}]{Nordhaus:2011aa}
{Nordhaus}, J., {Wellons}, S., {Spiegel}, D.~S., {Metzger}, B.~D., and
  {Blackman}, E.~G.: 2011,
\newblock {\em Proceedings of the National Academy of Science} {\bf 108}, 3135

\bibitem[\protect\astroncite{Ohlmann et~al.}{2016}]{Ohlmann2015}
Ohlmann, S.~T., R{\"{o}}pke, F.~K., Pakmor, R., and Springel, V.: 2016,
\newblock {\em The Astrophysical Journal} {\bf 816(1)}, L9

\bibitem[\protect\astroncite{Paczynski}{1976}]{Paczynski1976}
Paczynski, B.: 1976,
\newblock in P. Eggleton, S. Mitton, and J. Whelan (eds.), {\em Structure and
  Evolution of Close Binary Systems}, Vol.~73, pp 75--80, Dordrecht: Reidel,
  iau sympos edition

\bibitem[\protect\astroncite{Park and Bogdanovi{\'{c}}}{2017}]{Park2017}
Park, K. and Bogdanovi{\'{c}}, T.: 2017,
\newblock {\em The Astrophysical Journal} {\bf 838(2)}, 103

\bibitem[\protect\astroncite{Passy et~al.}{2012}]{Passy2012}
Passy, J.~C., {De Marco}, O., Fryer, C.~L., Herwig, F., Diehl, S., Oishi,
  J.~S., {Mac Low}, M.~M., Bryan, G.~L., and Rockefeller, G.: 2012,
\newblock {\em Astrophysical Journal} {\bf 744(1)}, 52

\bibitem[\protect\astroncite{{Paxton} et~al.}{2011}]{Paxton:2011aa}
{Paxton}, B., {Bildsten}, L., {Dotter}, A., {Herwig}, F., {Lesaffre}, P., and
  {Timmes}, F.: 2011,
\newblock {\em \apjs} {\bf 192}, 3

\bibitem[\protect\astroncite{{Paxton} et~al.}{2018}]{Paxton:2018aa}
{Paxton}, B., {Schwab}, J., {Bauer}, E.~B., {Bildsten}, L., {Blinnikov}, S.,
  {Duffell}, P., {Farmer}, R., {Goldberg}, J.~A., {Marchant}, P., {Sorokina},
  E., {Thoul}, A., {Townsend}, R.~H.~D., and {Timmes}, F.~X.: 2018,
\newblock {\em \apjs} {\bf 234}, 34

\bibitem[\protect\astroncite{Politano and Weiler}{2007}]{Politano2007}
Politano, M. and Weiler, K.~P.: 2007,
\newblock {\em The Astrophysical Journal} {\bf 665(1)}, 663

\bibitem[\protect\astroncite{{Reimers}}{1975}]{Reimers:1975aa}
{Reimers}, D.: 1975,
\newblock {\em Memoires of the Societe Royale des Sciences de Liege} {\bf 8},
  369

\bibitem[\protect\astroncite{Reyes-Ruiz and Lopez}{1999}]{ReyesRuiz1999}
Reyes-Ruiz, M. and Lopez, J.~A.: 1999,
\newblock {\em The Astrophysical Journal} {\bf 524(2)}, 952

\bibitem[\protect\astroncite{Ricker and Taam}{2012}]{Ricker2012}
Ricker, P.~M. and Taam, R.~E.: 2012,
\newblock {\em The Astrophysical Journal} {\bf 746(8pp)}, 74

\bibitem[\protect\astroncite{Sabach et~al.}{2017}]{Sabach2017}
Sabach, E., Hillel, S., Schreier, R., and Soker, N.: 2017,
\newblock {\em Monthly Notices of the Royal Astronomical Society} {\bf 472(4)},
  4361

\bibitem[\protect\astroncite{{Shappee} and {Thompson}}{2013}]{Shappee:2013aa}
{Shappee}, B.~J. and {Thompson}, T.~A.: 2013,
\newblock {\em \apj} {\bf 766}, 64

\bibitem[\protect\astroncite{{Soker}}{2015}]{Soker:2015aa}
{Soker}, N.: 2015,
\newblock {\em \apj} {\bf 800}, 114

\bibitem[\protect\astroncite{{Soker} et~al.}{2018}]{Soker:2018aa}
{Soker}, N., {Grichener}, A., and {Sabach}, E.: 2018,
\newblock {\em \apjl} {\bf 863}, L14

\bibitem[\protect\astroncite{{Thompson}}{2011}]{Thompson:2011aa}
{Thompson}, T.~A.: 2011,
\newblock {\em \apj} {\bf 741}, 82

\bibitem[\protect\astroncite{Toonen and Nelemans}{2013}]{Toonen2013}
Toonen, S. and Nelemans, G.: 2013,
\newblock {\em Astronomy {\&} Astrophysics} {\bf 557}, A87

\bibitem[\protect\astroncite{Tutukov and Yungelson}{1979}]{Tutukov1979}
Tutukov, A. and Yungelson, L.: 1979,
\newblock in P. Conti and C. de~Loore (eds.), {\em Mass Loss and Evolution of
  O-Type Stars}, pp 401--407, Reidel Publishing Company, Dordrecht, iau sympos
  edition

\bibitem[\protect\astroncite{{Villaver} and {Livio}}{2009}]{Villaver:2009aa}
{Villaver}, E. and {Livio}, M.: 2009,
\newblock {\em \apjl} {\bf 705}, L81

\bibitem[\protect\astroncite{Webbink}{1984}]{Webbink1984}
Webbink, R.~F.: 1984,
\newblock {\em The Astrophysical Journal} {\bf 277}, 355

\bibitem[\protect\astroncite{Zapolsky and Salpeter}{1969}]{Zapolsky1969}
Zapolsky, H.~S. and Salpeter, E.~E.: 1969,
\newblock {\em The Astrophysical Journal} {\bf 158}, 809

\bibitem[\protect\astroncite{Zorotovic et~al.}{2011}]{Zorotovic2011}
Zorotovic, M., Schreiber, M., G{\"{a}}nsicke, B., Rebassa-Mansergas, A., {Nebot
  G{\'{o}}mez-Mor{\'{a}}n}, A., Southworth, J., Schwope, A., Pyrzas, S.,
  Rodr{'}iguez-Gil, P., Schmidtobreick, L., Schwarz, R., Tappert, C., Toloza,
  O., and Vogt, N.: 2011,
\newblock {\em A{\&}A} {\bf 536}, L3

\bibitem[\protect\astroncite{Zorotovic et~al.}{2010}]{Zorotovic2010}
Zorotovic, M., Schreiber, M.~R., G{\"{a}}nsicke, B.~T., and {Nebot
  G{\'{o}}mez-Mor{\'{a}}n}, A.: 2010,
\newblock {\em A{\&}A} {\bf 520}, 86

\end{thebibliography}
\bibliographystyle{astron}

%%%%%%%%%%%%%%%%%%%%%%%%%%%%%%%%%%%%%%%%%%%%%%%%%%

\label{lastpage}
\end{document}